\documentclass[nofootinbib,prl,amsmath,twocolumn,amssymb,superscriptaddress,floatfix]{revtex4-2} %,linenumbers

\usepackage[english]{babel}
\makeatletter
\def\bbl@set@language#1{%
  \edef\languagename{%
    \ifnum\escapechar=\expandafter`\string#1\@empty
    \else\string#1\@empty\fi}%
  \@ifundefined{babel@language@alias@\languagename}{}{%
    \edef\languagename{\@nameuse{babel@language@alias@\languagename}}%
  }%
  \select@language{\languagename}%
  \expandafter\ifx\csname date\languagename\endcsname\relax\else
    \if@filesw
      \protected@write\@auxout{}{\string\select@language{\languagename}}%
      \bbl@for\bbl@tempa\BabelContentsFiles{%
        \addtocontents{\bbl@tempa}{\xstring\select@language{\languagename}}}%
      \bbl@usehooks{write}{}%
    \fi
  \fi}
\newcommand{\DeclareLanguageAlias}[2]{%
  \global\@namedef{babel@language@alias@#1}{#2}%
}
\makeatother
\DeclareLanguageAlias{en}{english}
\DeclareLanguageAlias{EN}{english}

\usepackage[colorlinks]{hyperref}
\usepackage{times}
\usepackage{graphicx}
\usepackage{amsmath, braket}
\usepackage{amssymb}
\usepackage{natbib}
\usepackage[normalem]{ulem}
\usepackage[dvipsnames]{xcolor}
\usepackage{soul}
\usepackage[version=4]{mhchem}
\usepackage[english]{babel}
\usepackage[parse-numbers=false]{siunitx}
\usepackage{xspace}
\usepackage{physics}
\usepackage{xr}

\newcommand{\nbse}{\ce{NbSe2}\xspace}

\newcommand{\musq}{$\SI{}{\micro m^2\,}$}
\newcommand{\musiq}{$\SI{}{\micro m^{-2}\,}$}
\newcommand{\panela}{\textbf{a}\xspace}
\newcommand{\panelb}{\textbf{b}\xspace}
\newcommand{\panelc}{\textbf{c}\xspace}
\newcommand{\paneld}{\textbf{d}\xspace}
\newcommand{\panele}{\textbf{e}\xspace}

\newcommand\blfootnote[1]{%
  \begingroup
  \renewcommand\thefootnote{}\footnote{#1}%
  \addtocounter{footnote}{-1}%
  \endgroup
}

\begin{document}

\title{Tunneling Spectroscopy of Two-Dimensional Materials Based on Via Contacts}

\author{Qingrui~Cao$^\dagger$}
\affiliation{Department of Physics, Carnegie Mellon University, Pittsburgh, PA 15213}
\author{Evan~J.~Telford$^\dagger$}
\affiliation{Department of Physics, Columbia University, New York, NY 10027}
\author{Avishai~Benyamini}
\affiliation{Department of Physics, Columbia University, New York, NY 10027}
\author{Ian~Kennedy}
\affiliation{Department of Physics, Columbia University, New York, NY 10027}
\author{Amirali~Zangiabadi}
\affiliation{Department of Applied Physics and Applied Mathematics, Columbia University, New York, NY 10027}
\author{Kenji~Watanabe}
\affiliation{Research Center for Functional Materials, National Institute for Materials Science, 1-1 Namiki, Tsukuba 305-0044, Japan}
\author{Takashi~Taniguchi}
\affiliation{International Center for Materials Nanoarchitectonics, National Institute for Materials Science, 1-1 Namiki, Tsukuba 305-0044, Japan}
\author{Cory~R.~Dean$^*$} %\email{cd2478@columbia.edu}
\affiliation{Department of Physics, Columbia University, New York, NY 10027}
\author{Benjamin~M.~Hunt$^*$} %\email{bmhunt@andrew.cmu.edu}
\affiliation{Department of Physics, Carnegie Mellon University, Pittsburgh, PA 15213}

\begin{abstract}
$$\bold{Abstract}$$
We introduce a novel planar tunneling architecture for van der Waals heterostructures based on via contacts, namely metallic contacts embedded into through-holes in hexagonal boron nitride (\textit{h}BN). We use the via-based tunneling method to study the single-particle density of states of two different two-dimensional (2D) materials, \nbse and graphene. In \nbse devices, we characterize the barrier strength and interface disorder for barrier thicknesses of 0, 1 and 2 layers of \textit{h}BN and study the dependence on tunnel-contact area down to $(44 \pm 14)^2 $ nm$^2$. For 0-layer \textit{h}BN devices, we demonstrate a crossover from diffusive to point contacts in the small-contact-area limit.  
In graphene, we show that reducing the tunnel barrier thickness and area can suppress effects due to phonon-assisted tunneling and defects in the \textit{h}BN barrier. This via-based architecture overcomes limitations of other planar tunneling designs and produces high-quality, ultra-clean tunneling structures from a variety of 2D materials.
\\
\\
$\bold{Keywords:}$ tunneling spectroscopy, via contacts, niobium diselenide, superconductivity, graphene, two-dimensional materials
\end{abstract}

\maketitle

\renewcommand{\figurename}{\textbf{Fig.}} % formatting for figure labels
\renewcommand{\thefigure}{\textbf{\arabic{figure}}}
\newcommand{\figtitle}[1]{\textbf{#1}\xspace}

\blfootnote{* cd2478@columbia.edu}
\blfootnote{* bmhunt@andrew.cmu.edu}

Tunneling spectroscopy is an indispensable experimental tool of modern condensed matter physics.  Vertical planar tunneling, which uses a fixed-width tunnel barrier, offers advantages over other spectroscopic tools such as scanning tunneling microscopy (STM).  One such advantage is the ability to tunnel in re-orientable and very large ($\geq$ 40 T) magnetic fields at dilution refrigerator temperatures ($\leq$ 30 mK), a capability that has application in, for example, determining the order parameter symmetry of novel two-dimensional superconductors \cite{sohn_unusual_2018}.
Another advantage is that the materials used for both the tunneling electrodes and the barriers themselves can be varied, allowing for a variety of tunneling devices to be implemented, including magnetic tunnel junctions and Josephson junctions \cite{zhu_magnetic_2006, makhlin_quantum-state_2001}.  The use of van der Waals materials in planar tunneling junctions offers the additional advantage that atomic precision of the tunneling distance can be achieved, and a variety of tunneling devices using van der Waals insulators, semiconductors, and magnetic insulators as tunneling barriers have been realized \cite{amet_tunneling_2012,britnell_electron_2012,singh_nanosecond_2016,gurram_bias_2017,bretheau_tunnelling_2017,khestanova_unusual_2018,dvir_spectroscopy_2018,song_giant_2018,keren_quantum-dot_2020,idzuchi_unconventional_2021}.
In the conventional design of planar tunneling junctions, however, it is difficult to control the area of a tunneling electrode that purely probes the bulk because such geometries necessarily include tunneling into the edge of a sample.  To perform spectroscopy using planar tunneling junctions, it is advantageous to have tunneling contacts with a small area to limit the effects of sample inhomogeneity, thereby approximating the tunneling from an STM tip or, in the case of a transparent barrier, to perform point-contact spectroscopy where the effective radius of the tunneling electrode is smaller than the electron mean-free path.

\begin{figure*}[ht!]
\begin{center}
\includegraphics[width=180mm]{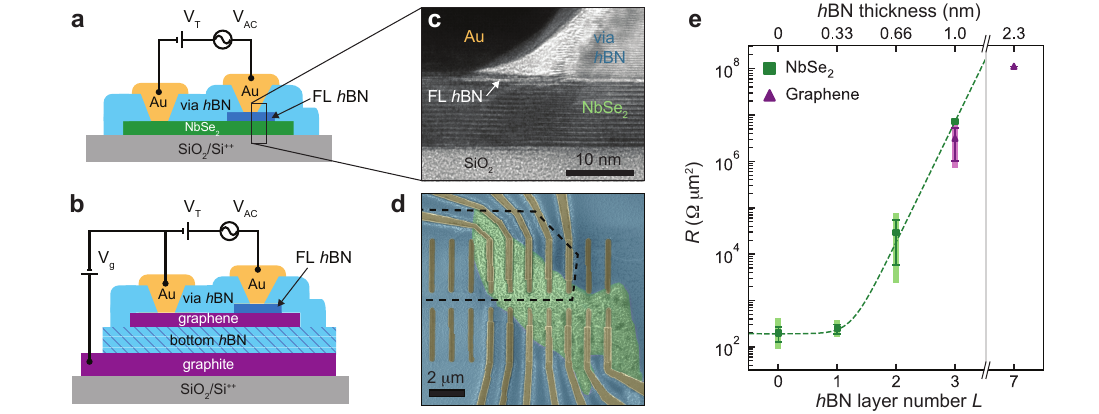}
\caption{\textbf{Characterization of via-based tunneling in van der Waals heterostructures.} \textbf{a} Schematic of a via-based NbSe$_2$ tunneling heterostructure.  \textbf{b} Schematic of a via-based monolayer graphene tunneling heterostructure.  The main difference from (\textbf{a}) is the ability to apply a local back gate voltage $V_g$.   \textbf{c} Cross-sectional TEM image of a Au/\textit{h}BN/\nbse junction. \textbf{d} Top-view false-color SEM micrograph of a representative \nbse tunneling device.  \nbse is green, $h$BN is blue, Au electrodes are yellow, and the black dashed line outlines the few-layer hBN tunnel barrier. \textbf{e} Measured tunneling resistance for \nbse (squares) and MLG (triangles) devices versus the number of \textit{h}BN layers $L$. Error bars represent the standard deviation of $\log_{10}(R)$ of all devices for a given $L$; shaded bars represent the full range of the data for that $L$. Dashed line is a fit to Eq.~(\ref{equation:transmission}). The anomalously low resistance for the 7L \textit{h}BN barrier in MLG is likely due to phonon-assisted tunneling in graphene (see main text). }
\label{fig:schematic}
\end{center}
\end{figure*}

In this work, we demonstrate planar tunneling junctions in van der Waals heterostructures whose size is limited in principle only by lithographic techniques.  
The junctions are based on metallic contacts that pass vertically through holes etched in exfoliated \textit{h}BN flakes, named ``via'' contacts in analogy with their function in conventional circuits \cite{telford_via_2018}.  Figs.~\ref{fig:schematic}\panela and \panelb depict side-view schematics of the two types of via-based tunneling structures studied. The "via" contacts are fabricated by reactive ion etching lithographically-patterned holes in 20-50 nm thick \textit{h}BN flakes followed by e-beam deposition of a non-sticking metal such as Au, Pd, or Pt inside the holes (see Supplemental Information for details) \cite{telford_via_2018}. The via-embedded \textit{h}BN flakes are then used subsequently to pick up a thinner (\textit{h}BN) tunnel barrier (typically $\leq 3$ layers thick) and then the target material using the dry-polymer transfer technique \cite{dean-1Dedge-2011}. 
This procedure can be performed entirely in an inert environment, ensuring good tunneling contact to air-sensitive materials.   
We study the dependence of via-based tunneling on the thickness of the \textit{h}BN tunnel barrier, on the tunneling area, and on the role of defects in the tunnel barrier.  This device geometry allows us to achieve junction areas as small as $(44 \pm 14)^2 $ nm$^2$, and in the limit of zero layers of \textit{h}BN, we demonstrate the crossover from a diffusive contact to a ballistic point contact as the tunneling area of the device is reduced. 

%%%%%%.   FIG. 1  %%%%%%%%%%
%%%%%%%%%%%%%%%%%%%%%%%%%%%%

We study two different target 2D materials ($<$ 25 nm thick \nbse and monolayer graphene, MLG), in order to investigate tunneling into a variety of electronic phases, such as into a 2D superconductor and into a quantum Hall insulator. \nbse is metallic and becomes superconducting below $T_c \approx$ 7 K.  MLG is a Dirac semimetal whose carrier density can be tuned by external gates and develops Landau levels (LLs) under perpendicular magnetic fields.  
In Fig.~\ref{fig:schematic}\panelc, we show a transmission electron microscopy (TEM) image of the vertical structure of a normal-metal/insulator/superconductor (NIS) junction. In the vicinity of the Au/\textit{h}BN/\nbse junction region, the interface is unbroken and atomically flat, indicating our fabrication procedure produces ultra-clean junction interfaces. Fig.~\ref{fig:schematic}\paneld shows a false-colored top-view scanning electron microscopy (SEM) image of a representative tunneling device.

The role of \textit{h}BN as a barrier can be approximated by calculating the transmission coefficient using a potential barrier with a finite height and width. The height in this approximation is determined by the offset of the \textit{h}BN conduction band and the Fermi energy of the tunneling electrode (determined by the alignment of their vacuum levels), and the width is set by the thickness of \textit{h}BN. The transmission coefficient $T$ then becomes
\begin{equation}
T = \left(1+\frac{V_0^2\sinh^{2} \left(t_{hBN}/\lambda\right)}{4E_F\left(eV_0-E_F\right)}\right)^{-1},
\label{equation:transmission}
\end{equation}
where $t_{hBN}$ is the \textit{h}BN thickness, $\lambda = \hbar / \sqrt{2m\left(eV_0-E_F\right)}$ is the characteristic wavelength, $E_F$ is the Fermi energy of the electrons in Au, and $eV_0$ is the barrier height. Fig.~\ref{fig:schematic}\panele shows a summary of the measured tunneling resistance normalized by tunnel area, $R$ (in $\Omega \cdot \SI{}{\micro m^2\,}$), as a function of the number of \textit{h}BN layers $L$.
Here, we compare the normal state tunneling resistance of the normal/insulating/normal (NIN) junctions in \nbse at 10 K (above $T_c$) and the NIN junctions in graphene doped to approximately $-2 \times 10^{12} \mathrm{\, cm^{-2}}$ at 4 K.
For $t_{hBN}/\lambda > 1$, the tunneling resistance versus \textit{h}BN layer number approximates an exponential dependence versus barrier width, consistent with previous measurements on Au/\textit{h}BN/Au tunnel junctions \cite{britnell_electron_2012}. Our data fits the predicted $\sinh^{2}$ dependence for $L \leq 3$, with an extracted length scale of $\lambda \approx 0.10$ nm. The corresponding barrier height relative to the Au Fermi energy, from $\lambda = \hbar / \sqrt{2m\left(eV_0-E_F\right)}$ and using the bare electron mass, is $eV_0-E_F = 3.82$ eV, consistent with Schottky barrier calculations on metal/\textit{h}BN interfaces \cite{bokdam_schottky_2014}. For thicker tunnel barriers ($L = 7$), the tunneling resistance is reduced due to phonon-assisted tunneling in MLG \cite{zhang_giant_2008}, as will be discussed in detail in later sections.

%%%%%%.   FIG. 2  %%%%%%%%%%
\begin{figure*}[ht!]
\begin{center}
\includegraphics[width=6.75in]{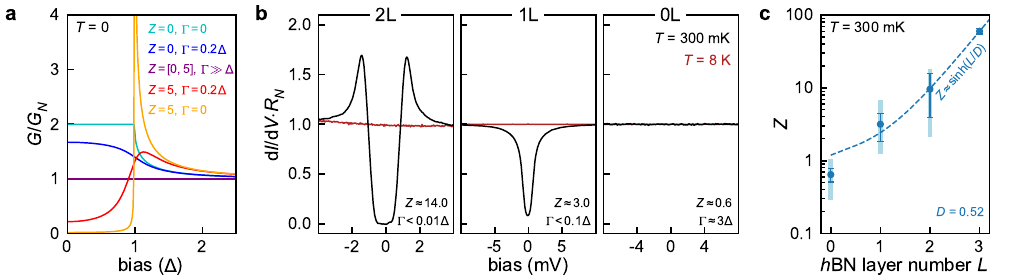}
\caption{\textbf{Via-based tunneling into \nbse}. \textbf{a} Predictions of BTK model for tunneling conductance $G$ normalized to normal-state conductance $G_N$, for representative combinations of barrier strength $Z$ and interface disorder $\Gamma$. \textbf{b} Measurements of differential conductance $G = dI/dV$ versus junction bias normalized to the normal-state resistance $R_N$ for $T<T_C$ (solid black lines) and $T>T_C$ (solid red lines) for 2, 1 and 0 layers of \textit{h}BN tunnel barriers, respectively. Extracted $Z$ and $\Gamma$ values are given in the insets. Results for 3L \textit{h}BN are qualitatively similar to those for 2L \textit{h}BN, and are shown in Fig.~\textbf{\ref{suppfig:3LhBNNbSe2}}. \textbf{c} Plot of barrier strength $Z$ versus \textit{h}BN layer number $L$. Dashed blue line is the fit to the data. The extracted fit parameter $D$ is given in the inset.}
\label{fig:nbse2_NIS}
\end{center}
\end{figure*}
%%%%%%%%%%%%%%%%%%%%%%%%%%%

We now turn to a detailed investigation of tunneling into the superconducting state of \nbse. 
In our analysis of the NIS tunneling, we employ the Blonder-Tinkham-Klapwijk (BTK) theory \cite{blonder_transition_1982}, in which the conductance of an NIS junction is primarily characterized by the barrier strength $Z\equiv H/\hbar v_F$, where $H$ models the strength of the delta-function potential at the normal metal-superconductor interface, i.e. $V(z) = H \delta (z)$, and $v_F$ is the Fermi velocity.  Large $Z$ corresponds to low-transparency junctions.  Fig.~\ref{fig:nbse2_NIS}\panela shows calculations of the tunnel conductance based on the BTK model for two tunneling limits, $Z=0$ and $Z=5$.  Also included in these calculations is the role of energy broadening of the density of states, which can come from interface disorder and inhomogeneity, external noise sources \cite{tamir_sensitivity_2019}, or inelastic scattering of electrons at the junction interface \cite{daghero_probing_2010}.  Although not in the original BTK derivation \cite{blonder_transition_1982}, it can be included by introducing a finite quasiparticle scattering lifetime $\hbar/\Gamma$ and substituting $E\rightarrow E + i\Gamma$. These two parameters, $Z$ and $\Gamma$, will be used to characterize via-based tunneling into \nbse through \textit{h}BN tunnel barriers of various thicknesses (see Supplemental Information for details on extracting $Z$ and $\Gamma$).

In Fig.~\ref{fig:nbse2_NIS}\panelb, we show the tunneling spectra of \nbse, acquired by varying the thickness of \textit{h}BN. In each panel we plot the differential conductance normalized to the normal-state resistance versus voltage bias above and below the superconducting transition temperature of \nbse. With $L=2$ layers of \textit{h}BN, we observe a canonical superconducting spectra for $T < T_C$ characterized by vanishing conductance below the superconducting gap, $\left|eV_T\right| < \Delta = 1.3$ meV, and distinct quasiparticle peaks appearing at the edge of the superconducting gap. 
For tunneling spectra acquired under similar conditions but with the tunnel barrier reduced to a single monolayer of \textit{h}BN ($L=1$), 
we now observe a finite conductance at zero bias and the quasiparticle peaks are almost non-existent. The observation of finite conductance below the gap is consistent with the expectation that decreasing layers of \textit{h}BN increases the transparency of the junction ($Z = 3.0$ compared to $Z = 14.0$ in the $L=2$ junction). The smeared-out quasiparticle peaks suggest an enhancement of interface broadening ($\Gamma < 0.1\Delta$ compared to $\Gamma < 0.01\Delta$ in the $L=2$ junction). This is confirmed in the junctions fabricated with no layers of \textit{h}BN.
The spectrum is completely bias independent as the interface broadening dominates the observed spectra ($\Gamma \approx 3\Delta$). This is not surprising considering our junctions are well outside the ballistic regime \cite{daghero_probing_2010} since the cross-sectional interface area is much larger than the mean free path of Au squared, $A \sim 1\,\musq \gg l^2_{mfp}$, where $A$ is the cross-sectional area of the Au/\nbse interface and $l_{mfp} \approx 37$ nm is the mean free path of the Au. This allows for inelastic scattering near the interface, broadening the energy of the incident electrons \cite{blonder_transition_1982, daghero_probing_2010}. In all spectra, for $T > T_C$, the differential conductance is independent of voltage bias, as expected for the normal state \cite{britnell_electron_2012}. In Fig.~\ref{fig:nbse2_NIS}\panelc, we summarize the extracted barrier strength $Z$ versus the number of \textit{h}BN layers $L$, where $Z$ was calculated assuming $\Gamma \approx 0$ (see Fig.~\textbf{\ref{suppfig:ZandGammacontours}} for details).
We find $Z \sim \sinh \left(L/D\right)$ well describes our data, with a fitting parameter $D$. The extracted value is $D = 0.52$, with a corresponding barrier height relative to the Au Fermi energy of $eV_0-E_F = 1.43$ eV.

%%%%%%.   FIG. 3 %%%%%%%%%%
\begin{figure}[ht!]
\begin{center}
\includegraphics[width=86mm]{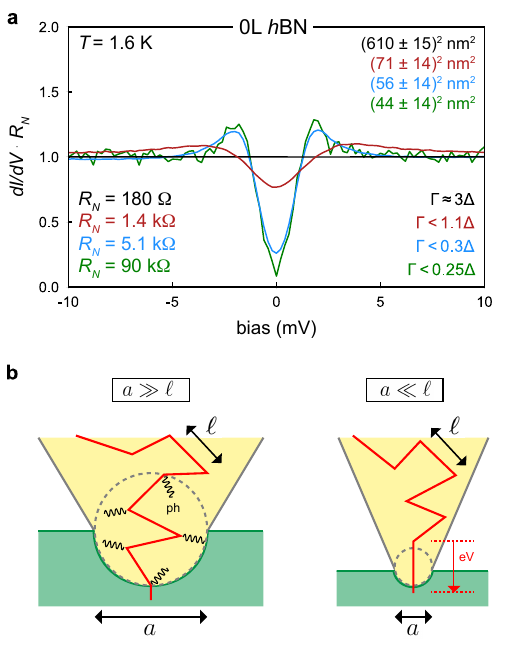}
\caption{\textbf{Crossover to point-contact spectroscopy in \nbse devices}. \textbf{a} Tunnel conductance versus junction bias normalized to the normal-state conductance for various contact cross-sectional areas for 0 layer \textit{h}BN tunnel barriers. The solid black line is the spectrum in Fig.~\ref{fig:nbse2_NIS}\panelb, given for reference. 
\textbf{b} Contributions to the current in a point contact \cite{daghero_probing_2010}.  (Left panel) In the ``thermal regime'' where $a=\sqrt{A} \gg l_{mfp}$, electrons undergo inelastic scattering in the contact region and transport through the junction is similar to normal transport between two conductors. (Right panel) In the ``ballistic regime'', electrons within $l_{mfp}$ are accelerated through the contact without scattering and gain energy $eV$, where $V$ is the applied voltage.  In this regime, it is possible to obtain spectroscopic information about the target material.}
\label{fig:areadep}
\end{center}
\end{figure}
%%%%%%%%%%%%%%%%%%%%%%%%%%%%

One interesting observation in the \textit{h}BN layer dependence is that adding a single layer of \textit{h}BN (with a low tunnel resistivity of $\sim 100 \, \Omega \cdot \SI{}{\um^2}$) decreases the energy broadening from $\Gamma \approx 3.0 \Delta$ to $\Gamma < 0.1 \Delta$. This implies that one could utilize monolayers of \textit{h}BN to engineer homogenous interfaces between disparate materials without significantly impacting the overall contact resistance. This effect has been observed in other systems \cite{cui_low-temperature_2017, kim_fermi_2017}, where it was demonstrated that adding single layers of \textit{h}BN between metal contacts and semiconductors prevents Fermi level
pinning and band bending due to interfacial defects. A similar effect occurs here in the NIS junctions, in which the \textit{h}BN barrier reduces inhomogeneity of the energy landscape caused by disorder and spatial inhomogeneities introduced during the fabrication process.

%%%%%%.   FIG. 4 %%%%%%%%%%
\begin{figure*}[ht!]
\begin{center}
\includegraphics[width=6.75in]{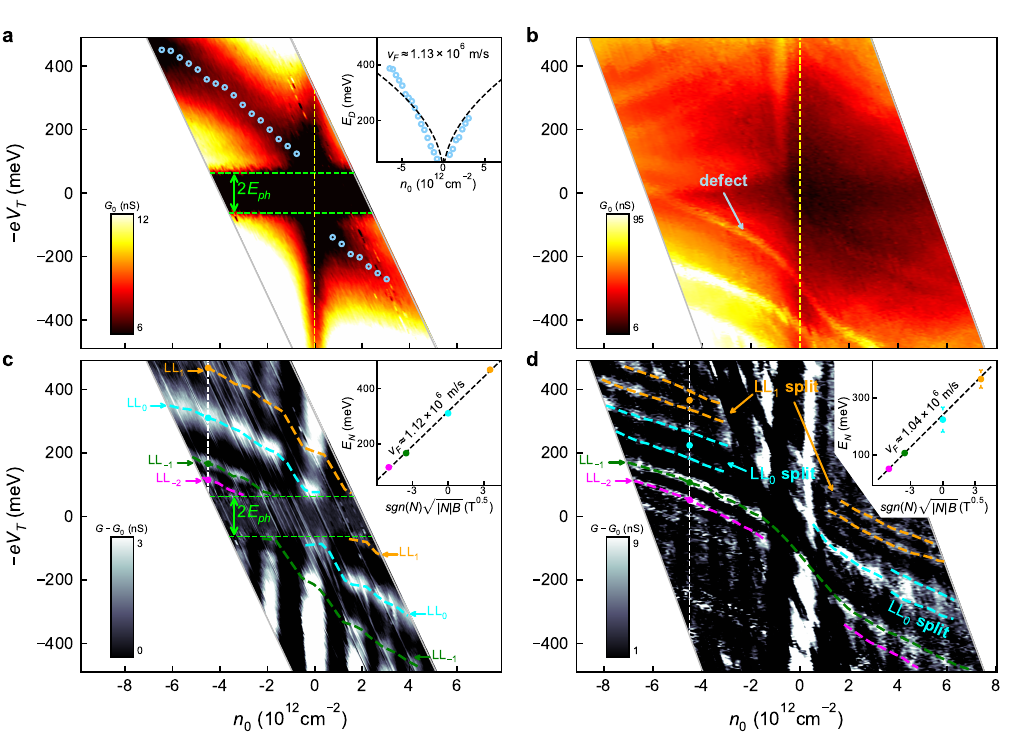}
\caption{\textbf{Via-based tunneling into monolayer graphene.} \textbf{a, b} Zero-field spectra of D1 ($L=7$, $A=3$ \musq) and D2 ($L=3$, $A=0.07$ \musq), respectively. 
\textbf{a} Light blue circles mark the locations of Dirac point. Light green dashed lines (also in panel \textbf{c}) mark the boundaries of the phonon gap, with $E_{ph} \approx 63$ meV. The inset shows the fit to Eq.~(\ref{equation:DiracPoint}), giving $v_F \approx 1.13 \times 10^{6}$ m/s, with data and extracted curve indicated by the light blue circles and black dashed curve, respectively.  Yellow dashed lines indicate the suppression of tunneling conductance $G_0$ near the charge neutrality point ($n_0 = 0$) due to the series resistance from in-plane transport.
\textbf{b} The light blue arrow marks the feature of a representative defect from \textit{h}BN tunnel barrier observed in the spectrum.
\textbf{c, d} $G-G_0$ map of D1 and D2 measured at 12.5 T, respectively. Trace of each $\mathrm{LL}_N$ ($N$ = -2, -1, 0 and 1) is indicated by unique color. The white dashed line is taken at $n_0' = -4.5\times10^{12}\,\mathrm{cm}^{-2}$.
\textbf{c} The inset shows the fit to Eq.~(\ref{equation:LLs}) at $n_0'$, giving $v_F \approx 1.12 \times 10^{6}$ m/s, with data and extracted line indicated by the dots with unique color and black dashed line, respectively.
\textbf{d} Arrows mark $\mathrm{LL}_0$ and $\mathrm{LL}_1$ spectral
features which show evidence of degeneracy lifting due to Zeeman splitting, with extracted $g \approx 113$ and $g \approx 80$, respectively, at $n_0' = -4.5\times10^{12}\,\mathrm{cm}^{-2}$.
The inset shows the fit to Eq.~(\ref{equation:LLs}) at $n_0'$ with the $\mathrm{LL}_{-1}$ and $\mathrm{LL}_{-2}$ points (indicated by pink and green dots), giving $v_F \approx 1.04 \times 10^{6}$ m/s. The black dashed line indicates the fitted line. The triangles for $\mathrm{LL}_0$ and $\mathrm{LL}_1$ are taken from the split peaks, and the dots are their corresponding midpoints.
}
\label{fig:mlg}
\end{center}
\end{figure*}
%%%%%%%%%%%%%%%%%%%%%%%%%%%%

For the relatively large-area 0L \textit{h}BN device shown in Fig.~\ref{fig:nbse2_NIS}\panelb ($\approx$ 0.6 \musq), the effect of interfacial disorder, within the BTK model, was significant: the spectrum showed negligible bias dependence and thus a complete absence of spectroscopic information.  Next, we investigate the role of contact area on tunneling spectra and present evidence that coherent tunneling spectra can be recovered by reducing the interfacial contact size (see Fig.~\ref{suppfig:small_via_dose} for fabrication details).  Fig.~\ref{fig:areadep}\panela shows tunneling spectra for four 0L \textit{h}BN devices with decreasing tunneling area, including the large-area device from Fig.~\ref{fig:nbse2_NIS}\panelb. As we reduce the contact size to $(71 \pm 14)^2 $ nm$^2$, we observe the emergence of distinct quasiparticle peaks and a suppression of conductance for $\left|eV_T\right| < \Delta$. Decreasing the contact size even further to $(56 \pm 14)^2 $ nm$^2$ and $(44 \pm 14)^2 $ nm$^2$, the quasiparticle peaks sharpen and the conductance suppression for $\left|eV_T\right| < \Delta$ deepens, indicating that $\Gamma$ decreases as the contact area decreases, down to $< 0.25 \Delta$ for $(44 \pm 14)^2 $ nm$^2$.  This behavior can be understood in the following way.  For large-area 0L \textit{h}BN, the tunneling spectra are dominated by interface inhomogeneities when the contact area ($A\sim 1$ \musq) is much greater than the squared mean free path of the Au contacts \cite{gall_electron_2016}, which results in qualitatively different behavior when $a=\sqrt{A} \gg \ell_{mfp}$ and when $a \lessapprox \ell_{mfp}$ (Fig.~\ref{fig:areadep}\panelb).  This indicates that we have crossed over from a diffusive contact (in the ``thermal'' regime) to a ballistic point contact \cite{daghero_probing_2010}.  A similar effect is observed for junctions fabricated with 1 layer of \textit{h}BN (see Fig.~\textbf{\ref{suppfig:1L_areadep}} for details). 
For the $L=1$ junctions, as the contact size decreases, the extracted $\Gamma$ values remain approximately constant while the quasiparticle peaks re-emerge, indicating lower broadening at the interface. We note that for small-area via contacts, there were sample-to-sample variations in junction resistance and barrier parameters $Z$ and $\Gamma$, which we suspect originated from uncontrolled variations in the individual fabrication processes.

Next, we turn to measurements of monolayer graphene (MLG) using the same method for producing via tunneling structures, and explore the \textit{h}BN thickness and contact area dependence on the spectra. We then use this platform to study Landau levels (LLs) at high magnetic field with various barrier thicknesses and junction sizes. 
Unlike \nbse, since MLG is a low-density semimetal, its carrier density $n_0$ can be considerably changed by external gate voltages:
\begin{equation}
n_0 = \left(C_g V_g + C_T V_T\right)/e,
\label{equation:carrierdensity}
\end{equation}
where $V_g$ is the back gate voltage, $V_T$ is the voltage applied to the via tunneling electrode, and $C_g$ and $C_T$ are the geometric capacitances per unit area between MLG and the back gate electrode, and between MLG and the via tunneling electrode, respectively.  Note that $V_T$ controls both the density $n_0$ via Eq.~(\ref{equation:carrierdensity}) as well as the energy at which the electrons tunnel, $E_T = -eV_T$.
Fig.~\ref{fig:mlg} shows results for the tunnel conductance in two representative MLG devices, D1 and D2, where the tunnel barrier thickness and contact area for D1 are 7 layers and 3\musq, respectively, and those for D2 are 3 layers and 0.07\musq. The results are plotted as a function of carrier density $n_0$ and tunneling energy $E_T$, where $n_0$ is determined by Eq.~(\ref{equation:carrierdensity}).  Spectroscopic features associated with the band structure of graphene appear curved in $(n_0, E_T)$ space \cite{amet_tunneling_2012}.
Since our measurements require the tunneling current to flow in the plane of MLG to the drain electrode \cite{dial_high-resolution_2007}, we observe additional, non-spectrosopic features due to this series transport resistance in both devices, which appear as vertical features in Fig.~\ref{fig:mlg}.

The spectrum of D1 at $B = 0$ (Fig.~\ref{fig:mlg}\panela) exhibits a prominent gap-like feature at low tunneling energy ($|E_T| \leq E_{ph} = 63$ meV), which arises from the intrinsic electron-phonon coupling in MLG due to the out-of-plane acoustic phonon modes near the $\mathbf{K}/\mathbf{K}'$ points \cite{mohr_phonon_2007} and has been reported in STM measurements \cite{zhang_giant_2008, yin_clarifying_2020}. 
Conductance is suppressed when $|E_T| \leq E_{ph}$ due to the thick tunnel barrier ($L=7$) and the absence of phonon-assisted tunneling, whereas above $E_{ph}$ it adds an additional conductance channel which appears as a sharp edge in the conductance maps.
Such phonon-assisted tunneling process enhances the amplitude of $G_0$ despite the usage of a thick tunnel barrier ($L=7$). 
Owing to energy loss during this inelastic scattering, locations of Dirac point (minima of $G_0$ in $E_T$ at given $n_0$ values, indicated by light blue circles), $V_D$, are offset by the phonon energy $E_{ph}$ from their true energy locations, $E_D$, giving
\begin{equation}
\left|eV_D\right| - E_{ph} = E_D = \hbar v_F \sqrt{\pi \left|n_0\right|},
\label{equation:DiracPoint}
\end{equation}
where $v_F$ is the Fermi velocity of MLG.
The inset in Fig.~\ref{fig:mlg}\panela shows the fit to Eq.~(\ref{equation:DiracPoint}), resulting in $v_F \approx 1.13 \times 10^6$ m/s, consistent with the previous transport and STM studies \cite{novoselov_two-dimensional_2005, zhang_giant_2008, chae_renormalization_2012}. Fig.~\ref{fig:mlg}\panelb shows the spectrum of D2 at B = 0. With a thinner tunnel barrier ($L=3$), we now observe the emergence of sharp peaks (indicated by the light blue arrow) in $G_0$ associated with single electron charging of the defects in \textit{h}BN, which is suppressed in D1 due to its larger barrier thickness, consistent with our conclusions from \nbse devices. The enhancement of tunneling current also leads to the disappearance of a clearly-defined phonon gap, as phonon-mediated tunneling no longer dominates in the presence of additional tunneling channels, similar to previous observations in STM \cite{yin_clarifying_2020}.

%%%%%%.   FIG. 5 %%%%%%%%%%
\begin{figure}[ht!]
\begin{center}
\includegraphics[width=3.375in]{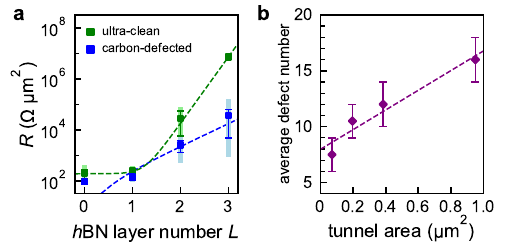}
\caption{\textbf{Role of defects in \textit{h}BN tunnel barriers.} \textbf{a} Normal-state resistance of \nbse devices fabricated with intentionally carbon-defected \textit{h}BN tunnel barriers (blue squares) and with ultra-clean \textit{h}BN tunnel barriers (green squares; data reproduced from Fig.~\ref{fig:schematic}\panele). \textbf{b} Number of observed defects versus tunnel area in four MLG devices fabricated with the same ultra-clean 3-layer \textit{h}BN tunnel barrier (a representative defect feature is shown for 0.07 \musq device in Fig.~\ref{fig:mlg}\panelb). Purple diamonds are obtained by estimating the number of distinct sharp peaks with the appropriate curvature in the $(n_0, E_T)$ space; full data sets for the four MLG devices are shown in Fig.~\textbf{\ref{suppfig:areadefects}}.}
\label{fig:defects_temp}
\end{center}
\end{figure}
%%%%%%%%%%%%%%%%%%%%%%%%%%%%

At finite magnetic fields, the DOS of graphene evolves into sharp peaks at energies of LLs, given by
\begin{equation}
E_N = E_D + v_F \cdot \mathrm{sgn}(N) \sqrt{2e\hbar |N|B},
\label{equation:LLs}
\end{equation}
where $N$ is the index for the $N$th LL ($\mathrm{LL}_N$), and $E_N$ is the energy of $\mathrm{LL}_N$. 
For clarity, we remove the global background that is independent of magnetic field by subtracting the differential conductance $G$ measured at $B = 12.5$ T from its counterpart $G_0$ measured at zero field, resulting in the $G - G_0$ maps shown in Figs.~\ref{fig:mlg}\panelc and \paneld for D1 and D2, respectively. Both spectra show distinct LLs, with each LL indicated by unique color in the maps. In Fig.~\ref{fig:mlg}\panelc, we identify the locations of $E_N$ ($N$ = -2, -1, 0 and 1) at $n_0' = -4.5\times10^{12}\,\mathrm{cm}^{-2}$ (white dashed line) and fit to Eq.~(\ref{equation:LLs}), resulting in $v_F \approx 1.12 \times 10^6$ m/s, very close to the value obtained from the zero-field spectrum in the previous section. In Fig.~\ref{fig:mlg}\paneld, each LL becomes more distinct and narrower in width in the $E_T$ direction, suggesting an improvement of the lifetime of Dirac quasiparticles due to the reduction in \textit{h}BN layer number $L$ \cite{miller_observing_2009, song_high-resolution_2010}. Additionally, we also notice several regions for $\mathrm{LL}_0$ and $\mathrm{LL}_1$ where the spectral features are split due to partial lifting of the fourfold spin-valley degeneracy. 
These energy splittings reach 82 meV and 58 meV at B = 12.5 T with $n_0' = -4.5\times10^{12}\,\mathrm{cm}^{-2}$
for $\mathrm{LL}_0$ and $\mathrm{LL}_{1}$, respectively, close to the previous graphene planar tunneling experiment probed by quantum dots \cite{keren_quantum-dot_2020}, where a spin-dependent origin (i.e. Zeeman splitting) was indicated. 
We further extract the $g$ factor using $\Delta E_Z = g \mu_B B$, where $\mu_B$ is the Bohr magneton, and find that 
$g \approx 113$ for $\mathrm{LL}_0$ and $g \approx 80$ for $\mathrm{LL}_1$, also consistent with the large $g$ values observed in \cite{keren_quantum-dot_2020}.

So far, we have examined how thickness and area of the tunnel barrier affect the zero-field spectra of \nbse and the zero- and finite-field spectra of monolayer graphene, all of which are possible because of the via contact geometry. Lastly, we discuss the role of defects on the overall tunneling spectra. This has been extensively studied in NIN vdW junctions using \textit{h}BN as the tunnel barrier \cite{liu_defects_2018, amet_tunneling_2012, chandni_evidence_2015, greenaway_tunnel_2018}.
For \nbse devices, we repeat our characterization of
normal-state junction resistance versus tunnel barrier thickness with intentionally carbon-defected \textit{h}BN \cite{dirty_hbn_synthesis} (see Supplemental Information for details) and compare with the results obtained from pristine \textit{h}BN (Fig.~\ref{fig:defects_temp}\panela). The pristine \textit{h}BN has a defect density of $<10^{18}$ cm$^{-3}$, whereas the carbon-defected \textit{h}BN has a carbon defect density of $ 10^{18}-10^{20}$ cm$^{-3}$ \cite{chandni_evidence_2015,onodera_carbon-rich_2019,dirty_hbn_synthesis}. The role of carbon defects is to effectively reduce the intrinsic \textit{h}BN band gap by contributing mid-gap states \cite{huang_defect_2012, wang_local_2016,dirty_hbn_synthesis}, which is reflected in the normal-state resistance data.
The thickness dependence is still well-described by a $\sinh^{2}$ dependence in Eq.~(\ref{equation:transmission}), but the extracted characteristic length becomes $\lambda \approx 0.29$ nm, three times as much as the pristine value and the corresponding barrier height relative to the Au Fermi energy is $eV_0-E_F = 0.44$ eV, indicating the presence of mid-gap states in the \textit{h}BN band structure due to carbon defects. 
With defect densities as low as $<10^{18}$ cm$^{-3}$, pristine \textit{h}BN serves as an excellent platform for studying individual defect states as NIN junctions can be fabricated with cross-sectional areas $A < 10$\musq. Our via-based method further allows us to precisely control the area of the contact sizes, and therefore to study the dependence of the number of observed defects on the contact size in MLG devices. Fig.~\ref{fig:defects_temp}\panelb shows the number of defects observed in tunneling spectra with different contact areas using a 3-layer \textit{h}BN as the tunnel barrier (see full spectra in Fig.~\textbf{\ref{suppfig:areadefects}}).  A representative defect feature is labelled in Fig.~\ref{fig:mlg}\panelb for the $A = 265^2$ nm$^2$ tunnel barrier.   We find that the defect numbers grow linearly with increasing tunnel area, giving about 17\musiq. The low defect density compared with the previously reported value \cite{chandni_evidence_2015} is likely due to the suppression of defect features in the spectra as well as the limited energy window in which they can be observed. 

In summary, by studying the spectra of \nbse and MLG at low temperature and high magnetic fields with various tunnel junction areas and $h$BN barrier thicknesses, we demonstrate that the via-based planar tunneling method has multiple advantages over conventional planar tunneling geometries in vdW heterostructures and over STM. The via platform avoids exposing any interface to air or lithographic polymers and permits tunneling areas that are limited, in principle, only by lithographic techniques.  We found that for $\sim$1\musq-area tunnel barriers, 2L $h$BN optimizes the combination of low interface transparency $Z$ and low interface disorder $\Gamma$ for tunneling into 2D superconductors; 2L $h$BN has also been found to be the ideal thickness as a tunnel barrier for spin injection in graphene-based spintronic devices \cite{singh_nanosecond_2016,gurram_bias_2017}.  In the limit of small tunneling area (down to $(44 \pm 14)^2 $ nm$^2$) and 0L $h$BN, we found that point-contact spectra could be realized.  This suggests that via-based tunneling devices could be a powerful probe of the spin structure of triplet or mixed singlet-triplet superconductors by studying the tunneling into the same sample from multiple via point contacts having different degrees of spin polarization \cite{chen_unified_2012}.  Such high quality tunneling structures could also be applied to other low-dimensional systems, such as superconducting contacts to semiconductors \cite{ramezani_superconducting_2021}, superconducting-ferromagnetic-superconducting (SFS) junctions, or high-$T_c$ superconductors, or combined with other techniques such as the capacitive detection of vertical planar tunneling currents \cite{dial_high-resolution_2007} to reduce the series transport resistance in semiconducting/insulating devices.

\section{Supporting Information}
Identification and contrast calibration of ultra-thin $h$BN; device fabrication details; measurement setup; BTK calculation of \nbse tunneling spectra; simulation of graphene tunneling spectra; extended transport and tunneling data.

\section{Author Contributions}

Q.C. and E.J.T. contributed equally to this work. Q.C. fabricated, measured, and analyzed data from the graphene-based tunneling devices. E.J.T., A.B., and I.K. fabricated, measured, and analyzed data from NbSe\textsubscript{2}-based tunneling devices. E.J.T. performed the scanning electron microscopy. A.Z. performed the transmission electron microscopy. K.W. and T.T. provided \textit{h}BN and carbon-defected \textit{h}BN crystals. C.R.D. and B.M.H. supervised the research. All authors contributed to writing the manuscript.

\section{Acknowledgements}
We acknowledge stimulating conversations with Sergio de la Barrera. B.M.H. and Q.C. acknowledge support from the Department of Energy under the Early Career award program (DE-SC0018115) for device measurement, data analysis, and preparation of the manuscript. Q.C. was supported by the Office of Naval Research (N-000142112443) for device fabrication.  C.R.D., E.J.T., A.B., and I.K. were supported by the Columbia MRSEC on Precision-Assembled Quantum Materials (PAQM) - DMR-2011738 - for device fabrication, tunneling spectroscopy, and preparation of the manuscript.

% \clearpage
\bibliography{main_SI_combined.bib}
\newpage
\onecolumngrid % comment to retain two-column format
\cleardoublepage
\renewcommand\thesection{S\arabic{section}}
\renewcommand\thesubsection{S\thesection.\arabic{subsection}}
\renewcommand{\thefigure}{\textbf{S\arabic{figure}}}
\renewcommand{\theequation}{S\arabic{equation}}
\section{\texorpdfstring{Supplementary Information \\ Tunneling Spectroscopy of Two-Dimensional Materials Based on Via Contacts}{Supplementary Information for Tunneling Spectroscopy of Two-Dimensional Materials Based on Via Contacts}}

\subsection{Identification and Contrast Calibration of Ultra-thin \textit{h}BN}
\begin{figure*}[ht!]
\begin{center}
\includegraphics[width=7in]{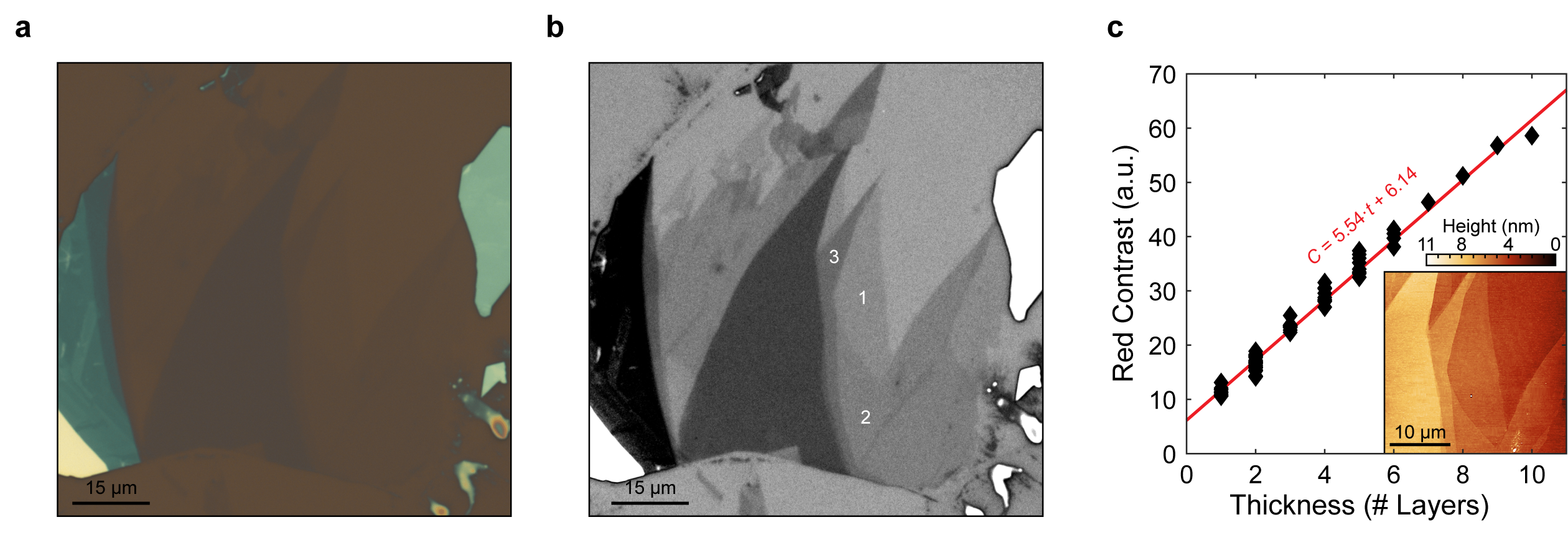}
\caption[Identification and characterization of few-layer \textit{h}BN]{{\bf Identification and characterization of few-layer \textit{h}BN.} {\bf a},{\bf b}) Unprocessed ({\bf a}) and processed ({\bf b}) optical images of mono, bi, and trilayer \textit{h}BN flakes on a $90$ nm SiO\textsubscript{2}/Si\textsuperscript{++} substrate. In ({\bf b}), the mono, bi, and trilayers of \textit{h}BN are labeled by their corresponding layer number. {\bf c}) Red optical contrast versus \textit{h}BN thickness. A linear fit to the data is given by the solid red line. The fit parameters are given in the inset. Lower right inset: atomic force microscope topography of the flake in ({\bf a}) and ({\bf b}).}
\label{suppfig:thinhBN}
\end{center}
\end{figure*}

To quickly and reliably identify the thickness of ultra-thin \textit{h}BN flakes ($<$10 layers), we developed a contrast calibration curve for flakes exfoliated on 90 nm SiO\textsubscript{2}/Si\textsuperscript{++} substrates.  Substrates with 90 nm SiO\textsubscript{2}/Si\textsuperscript{++} were chosen over 285 nm SiO\textsubscript{2}/Si\textsuperscript{++} substrates due to the increased contrast for few-layer \textit{h}BN \cite{gorbachev-geim-blake-2008}. First, a series of images of \textit{h}BN flakes with varying thicknesses was collected using a Nikon Eclipse LV150N microscope fitted with a Nikon DS-Fi3 camera (\textbf{Figure \ref{suppfig:thinhBN} a}). The images were then shading corrected and the flake contrast was extracted using Gwyddion (\textbf{Figure \ref{suppfig:thinhBN} b,c}). We found that red color contrast was the most significant. The contrast was then correlated to flake thickness (\textbf{Figure \ref{suppfig:thinhBN} c}), which was measured through atomic force microscopy and tunneling resistance \cite{liam-novoselov-2012}.

Atomic force microscopy was performed in a Bruker Dimension Icon\textsuperscript{\textregistered} using OTESPA-R3 tips in tapping mode. Flake thicknesses were extracted using Gwyddion to measure histograms of the height difference between the substrate and the desired \textit{h}BN flake.

\subsection{Fabrication of Au/\textit{h}BN/NbSe\textsubscript{2} Tunneling Devices}
Tunneling devices were fabricated from NbSe\textsubscript{2} flakes using the via contact technique \cite{telford-nl-2018} in which \textit{h}BN with embedded Au electrodes was used to pick up ultra-thin flakes of \textit{h}BN (ranging in thickness from 1-3 layers) and subsequently placed onto the desired NbSe\textsubscript{2} flake using the dry-polymer-transfer technique \cite{dean-1Dedge-2011}. The via contacts and \textit{h}BN tunnel barrier were prepared under ambient conditions, whereas the NbSe\textsubscript{2} flakes were prepared under inert conditions in an N\textsubscript{2} glovebox with $<$5 ppm O\textsubscript{2} and $<$0.5 ppm H\textsubscript{2}O. All exfoliated flakes were prepared using mechanical exfoliation with Scotch\textsuperscript{\textregistered} Magic{\texttrademark} tape \cite{field-effect-2004, peter-huang-2015}. First, the via contacts and \textit{h}BN tunnel barrier were picked up in air then transferred into a glovebox where the heterostructure was placed onto the desired NbSe\textsubscript{2} flake under inert conditions. Bonding pads (Cr + Au: 2nm + 80nm) were then designed and deposited using conventional electron-beam lithography and deposition techniques. All devices were diced by hand and bonded to a 16-pin DIP socket for measurement in cryogenic systems. Between fabrication steps, all devices were stored in the N\textsubscript{2} glovebox to avoid sample degradation.

\subsection{Tunneling Measurements on Au/\textit{h}BN/NbSe\textsubscript{2} Tunnel Junctions}
Tunnel junction resistance was measured in a 2-terminal differential configuration whereby a small AC voltage bias was superimposed on top of a DC voltage bias. The corresponding AC current was measured as a function of applied DC voltage, temperature, and magnetic field. For all data presented in the manuscript, the tunnel conductance is defined as $G = \frac{dI}{dV} (V_{DC}) = \frac{I_{AC}}{V_{AC}} (V_{DC})$, where $V_{AC}$ and $I_{AC}$ are the AC voltage bias and current and $V_{DC}$ is the applied DC voltage bias. The normal state resistance is defined as the junction resistance when the DC voltage bias is greater than the superconducting gap, $R_N = \frac{1}{dI/dV} (eV_{DC}> \Delta_{NbSe2})$. For all data presented in the manuscript, $V_{AC}$ was set to $<$50 $\mu$V. The AC voltage bias was applied using an SRS830 lock-in amplifier with a reference frequency of 17.777Hz and the AC current was measured using the same lock-in amplifier. The DC voltage bias was applied using a Keithley 2400. The AC and DC voltage sources were connected in parallel to a voltage divider, the output of which was directly connected to the sample (\textbf{Figure \ref{suppfig:measurementschematic}}). To ensure the DC voltage was dropped predominantly across the relevant Au/\textit{h}BN/NbSe\textsubscript{2} interface, the DC+AC voltage source was injected at the tunnel barrier and drained from low-resistance Au/NbSe\textsubscript{2} contacts (\textbf{Figure \ref{suppfig:measurementschematic}}). For measurements of low-resistance tunnel junctions (with $<$2 layers of \textit{h}BN) multiple Au/NbSe\textsubscript{2} contacts were used as the drain to reduce the total drain contact resistance. Measurements were performed either in a $^3$He cryostat with temperatures ranging from 300 mK up to 10 K or a pumped $^4$He cryostat with temperatures ranging from 1.6 K up to 10 K.

\subsection{Synthesis of Carbon-defected \textit{h}BN}
\begin{figure*}[ht!]
\begin{center}
\includegraphics[width=7in]{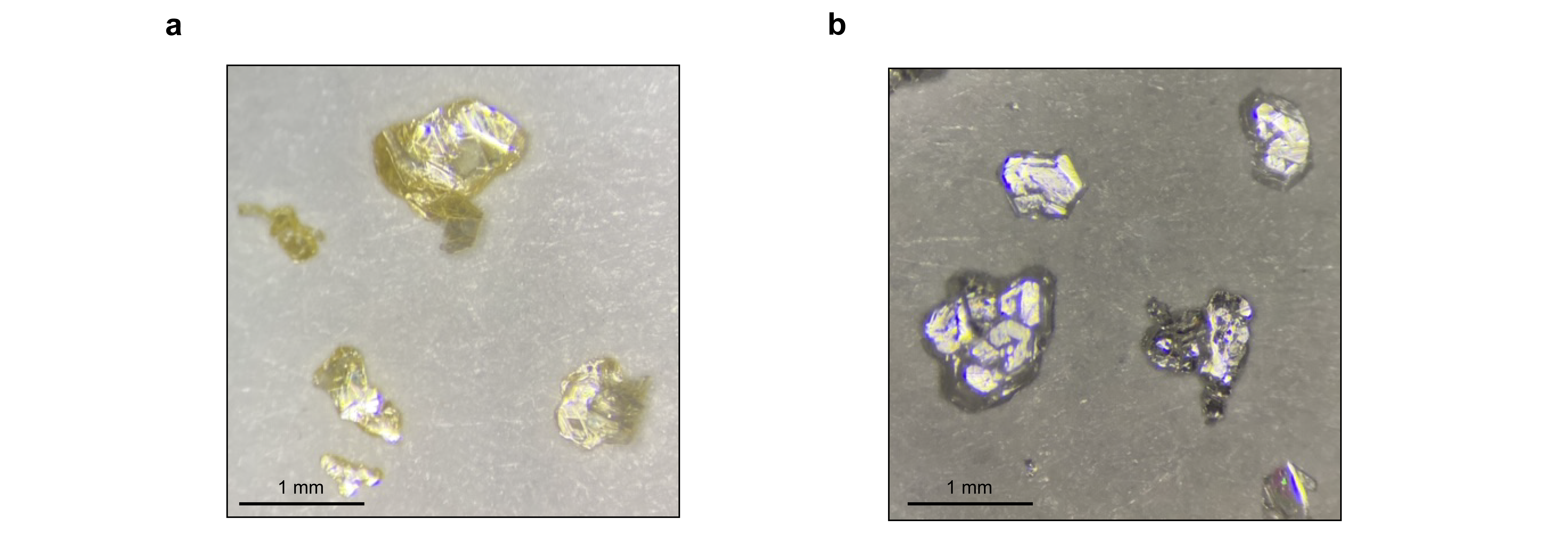}
\caption[Optical images of bulk \textit{h}BN crystals]{{\bf Optical images of bulk \textit{h}BN crystals.} Optical images of carbon-defected (\textbf{a}) and ultra-pure (\textbf{b}) bulk \textit{h}BN crystals. Synthesis and characterization of pristine and carbon-defected \textit{h}BN are outlined in references \cite{dirty_hbn_synthesis,dirty_hBN}.}
\label{suppfig:cleandirtyhbn}
\end{center}
\end{figure*}

Carbon-defected \textit{h}BN crystals were obtained by post treatment of ultra-clean \textit{h}BN crystals by annealing with graphite powder using the procedure outlined in reference \cite{dirty_hbn_synthesis}. As carbon diffusion takes place, the color of the \textit{h}BN crystals changes from white to yellow (\textbf{Figure \ref{suppfig:cleandirtyhbn})}.

\subsection{Scanning Electron Microscopy}
Scanning electron micrographs were collected on a Zeiss Sigma VP scanning electron microscope (SEM) using a beam energy of 5 kV.

\subsection{Transmission Electron Microscopy}
Thermo Scientific Helios NanoLab 660 (equipped with focused ion-beam) was used to prepare thin foils for transmission electron microscopy (TEM). In order to protect the surface against the ion-milling process, amorphous platinum (2$\SI{}{\micro m}$ thick) was sputtered on the top surface by the electron and ion-beam, respectively. High-resolution TEM images were acquired by Thermo Scientific Talos F200X (S)TEM at an accelerating voltage of 200 kV using a 100$\SI{}{\micro m}$ objective aperture.

\subsection{Fabrication of Au/\textit{h}BN/graphene Tunneling Devices}
Tunneling devices were fabricated from graphene using the via contact technique where \textit{h}BN was embedded with Au or Pd electrodes that were arranged in two rows. First, the via contact \textit{h}BN flakes were picked up using the dry-polymer-transfer technique \cite{dean-1Dedge-2011}. Thin \textit{h}BN tunnel barriers were subsequently picked up such that the few-layer \textit{h}BN was located under only one of the rows of vias, after which we picked up the graphene flake as well as the remaining \textit{h}BN capping layer (20-50 nm thick) and the bottom graphite gate ($>$30 nm thick). The entire heterostructure was then transferred onto a SiO\textsubscript{2}/Si\textsuperscript{++} substrate. One row of the via contacts served as tunneling electrodes with precisely defined and located tunneling areas, and the other row of via contacts made direct contact to the graphene and served as drain electrodes. All exfoliated flakes were prepared using mechanical exfoliation with Scotch\textsuperscript{\textregistered} Magic{\texttrademark} tape \cite{field-effect-2004, peter-huang-2015}. As a final step, bonding pads (Cr + Pd + Au: 2nm + 40nm + 50nm) were designed and deposited using standard electron-beam lithography and deposition. All fabrication steps were performed under ambient conditions.

\subsection{Tunneling Measurements on Au/\textit{h}BN/graphene Tunnel Junctions}
\begin{figure*}[ht!]
\begin{center}
\includegraphics[width=7in]{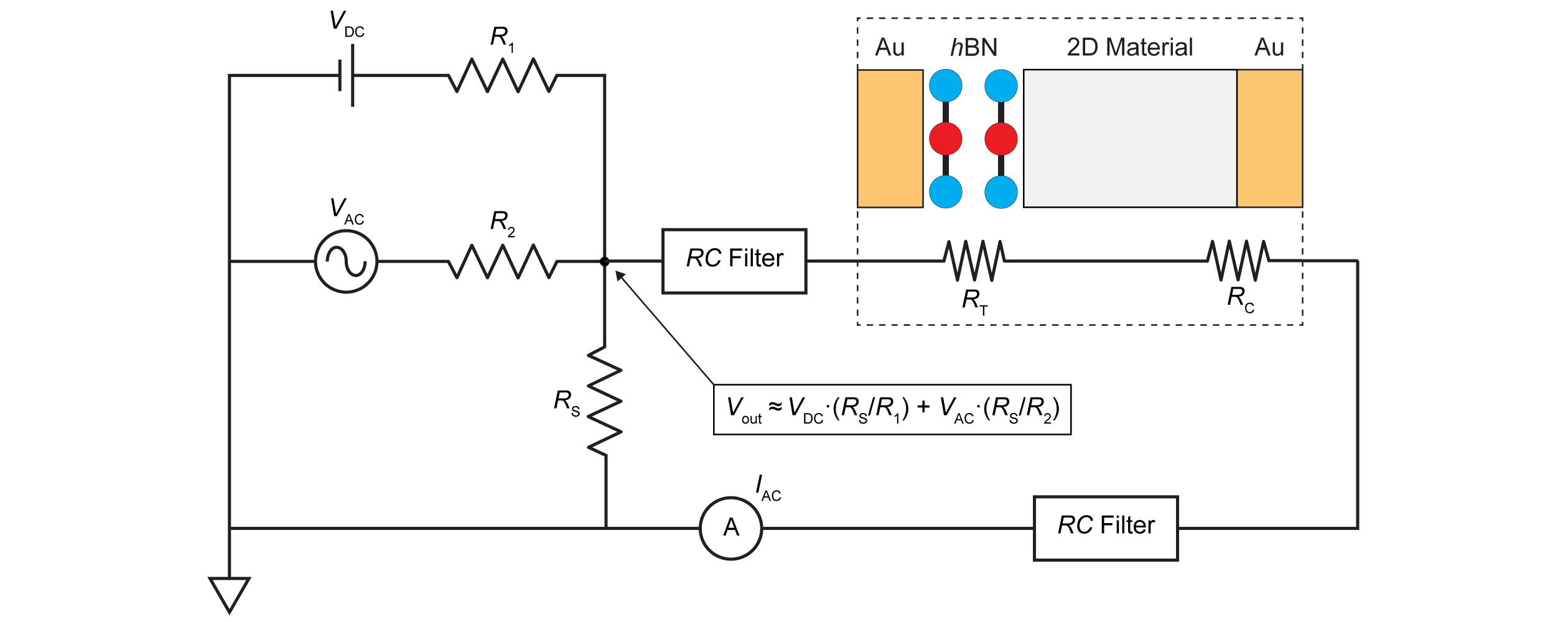}
\caption[Schematic of the tunneling measurement setup]{{\bf Schematic of the tunneling measurement setup.} For the Au/\textit{h}BN/NbSe\textsubscript{2} experiments, $R_1 = 100\,k\Omega$, $R_2 = 1\,M\Omega$, $R_S = 100\,\Omega$. The {\em RC} filters are homemade low-pass filters with 3 dB attenuation at $\sim$5 kHz \cite{Benyamini2019, Idan_filterting_sciadv}. For the Au/\textit{h}BN/graphene experiment, $R_1 = R_2 = 10\,k\Omega$, $R_S = 100\,\Omega$. The $RC$ filters are also low-pass filters with 3 dB attenuation at $\sim$5 kHz. The dashed square represents the fabricated tunnel junction devices.}
\label{suppfig:measurementschematic}
\end{center}
\end{figure*}

\begin{figure*}[ht!]
\begin{center}
\includegraphics[width=6.75in]{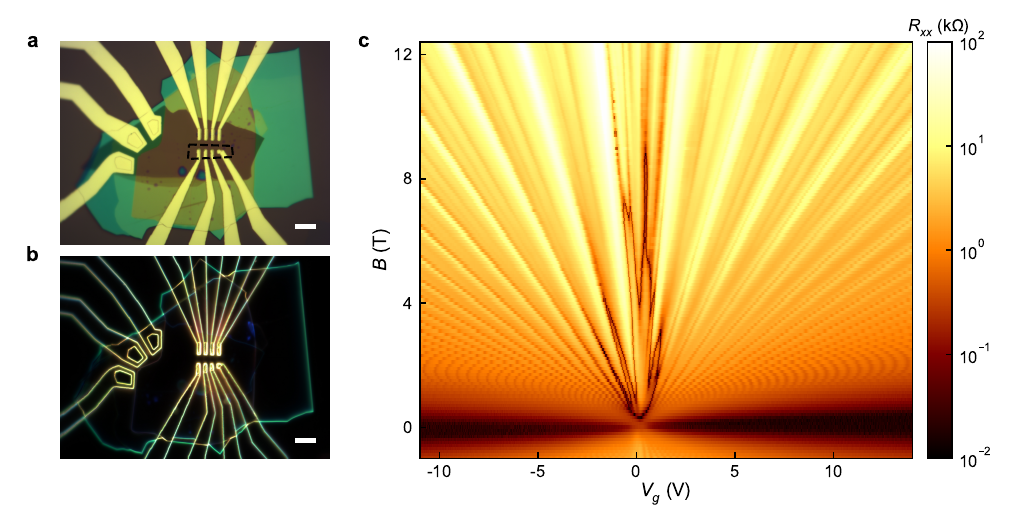}
\caption{\textbf{Characterization of a representative monolayer graphene device.} \textbf{a, b} Microscopic image of a MLG device with 100x magnification under bright field (\textbf{a}) and dark field (\textbf{b}). The black dashed line in (\textbf{a}) outlines the \textit{h}BN tunnel barrier. Both scale bars are 5$\SI{}{\micro m}$. \textbf{c} Four-terminal resistance measurement (without \textit{h}BN tunnel barrier) as a function of back gate voltage and magnetic field.}
\label{suppfig:transport}
\end{center}
\end{figure*}

Tunnel junction resistance was measured in a 2-terminal differential configuration whereby a small AC voltage bias was superimposed on top of a DC voltage bias. The corresponding AC current was measured as a function of applied DC voltage, temperature, and magnetic field. For all data presented in the manuscript, the tunnel conductance is defined as $G = \frac{dI}{dV} (V_{DC}) = \frac{I_{AC}}{V_{AC}} (V_{DC})$, where $V_{AC}$ and $I_{AC}$ are the AC voltage bias and current and $V_{DC}$ is the applied DC voltage bias. For all data presented in the manuscript, $V_{AC}$ was set between 5-10 mV. The AC voltage bias was applied using an SRS860 lock-in amplifier with a reference frequency of 13Hz and the AC current was measured using the same lock-in amplifier. An SRS570 preamplifier was placed in series before the current-measuring lockin to reduce the noise (lowpass filter, 12db, cutoff 1kHz). The DC voltage bias was applied using a Keithley 2400. The AC and DC voltage sources were connected in parallel to a voltage divider, the output of which was directly connected to the sample (\textbf{Figure \ref{suppfig:measurementschematic}}). To ensure the DC voltage was dropped predominantly across the relevant Au/\textit{h}BN/graphene interface, the DC+AC voltage source was injected at the tunnel barrier and drained from low-resistance Au/graphene contacts (\textbf{Figure \ref{suppfig:measurementschematic}}). The electrostatic gate voltage was applied using a Yokogawa GS200 DC voltage source. All measurements were performed in a dilution refrigerator with temperature ranging from 35 mK to 4.5 K. \textbf{Figure \ref{suppfig:transport}} shows the microscope images and transport characterization of a representative MLG device.

\subsection{Blonder-Tinkham-Klapwijk (BTK) Calculations of NbSe\textsubscript{2} Tunneling Spectra}
\subsubsection{Overview of BTK Theory Calculations}
Using BTK theory \cite{PhysRevB.25.4515}\cite{Sohn-NbSe2-2018}, we can directly model the tunneling conductance versus $V_{DC}$ with various barrier strengths $Z$ and inhomogeneity parameters $\Gamma$. The tunneling rate $I_{NS}$ of the normal-insulator-superconductor (NIS) junction as a function of $V_{DC}$ and temperature $T$ can be written as
\begin{equation}
I_{NS}(V_{DC}) \propto \int_{-\infty}^{\infty} \left (f(E-eV_{DC},T) - f(E,T)\right ) \left [1+AR(E)-R(E) \right ]dE
\end{equation}
where $f$ is the Fermi function and $AR(E)$ and $R(E)$ are the probabilities for Andreev reflection and normal reflection, respectively. For comparison to experiments, we normalize the NIS current by the magnitude of the current in the normal state (i.e. assuming both materials were normal metals), which is found by taking $I_{NS}$ in the limit $AR(E) \to 0$ and $\Delta \to 0$, $I_{NN}(V) \propto \frac{V}{1+Z^2}$. Our final expression for interface current is the following
\begin{equation}
\frac{I_{NS}(V_{DC})}{I_{NN}} =\frac{1+Z^2}{eV_{DC}} \int_{-\infty}^{\infty} \left (f(E-eV_{DC},T) - f(E,T)\right ) \left [1+AR(E)-R(E) \right ]dE
\end{equation}
where $E$ is the electron energy, $V_{DC}$ is the applied DC voltage, $e$ is the electron charge, $T$ is the sample temperature, $AR(E)$ and $R(E)$ are the Andreev reflection and normal electron reflection probabilities, respectively, and $f$ is the Fermi function. Ordinary reflection of electrons reduces the overall tunneling current, but Andreev reflections enhance it by transmitting a Cooper pair for each reflected hole. $AR(E)$ and $R(E)$ can be determined by matching wave-function and wave-function-derivative boundary conditions on either side of the junction, assuming the gap $\Delta$ is zero on the metal side and constant on the superconductor side. The interface potential is modeled as a delta function $V(z) = H\delta(z)$ with a dimensionless barrier strength parameter $Z = \frac{k_F}{2\epsilon_F} H$, where $k_F$ and $\epsilon_F$ are the Fermi wave vector and energy, respectively. The total electron transmission probability can be written as
\begin{equation}
    1 + AR(E) - R(E) = \tau\frac{1 + \tau|\gamma(E)|^2 + (\tau-1)|\gamma(E)^2|^2}{|1 + (\tau-1)\gamma(E)^2|^2}
\end{equation}
where $\tau = \frac{1}{1 + Z^2}$ is the transparency of the junction and
\begin{equation}
    \gamma(E) = \frac{N_q(E)-1}{N_p(E)}, N_q(E) = \frac{E}{\sqrt{E^2 - \Delta^2}}, N_p(E) = \frac{\Delta}{\sqrt{E^2 - \Delta^2}}
\end{equation}
In our experiments, when we extract quantitative information about the tunnel barrier, the sample temperature is much lower than the superconducting gap of NbSe\textsubscript{2}, so we can take the limit of $T \to 0$, simplifying the expression for barrier conductance
\begin{equation}
\frac{G_{NS}}{G_{NN}} =\frac{1}{\tau} \frac{d}{d(eV_{DC})} \int_{0}^{eV_{DC}} \left [1+AR(E)-R(E) \right ]dE
\end{equation}
To account for the effects of energy broadening and inhomogeneity, the model can be modified by replacing $E \to E + i\Gamma$ where $\frac{1}{\Gamma}$ is interpreted as a finite quasiparticle life time.

\subsubsection{Extracting $Z$ and $\Gamma$ from Tunneling Spectra}
\begin{figure*}[ht!]
\begin{center}
\includegraphics[width=7in]{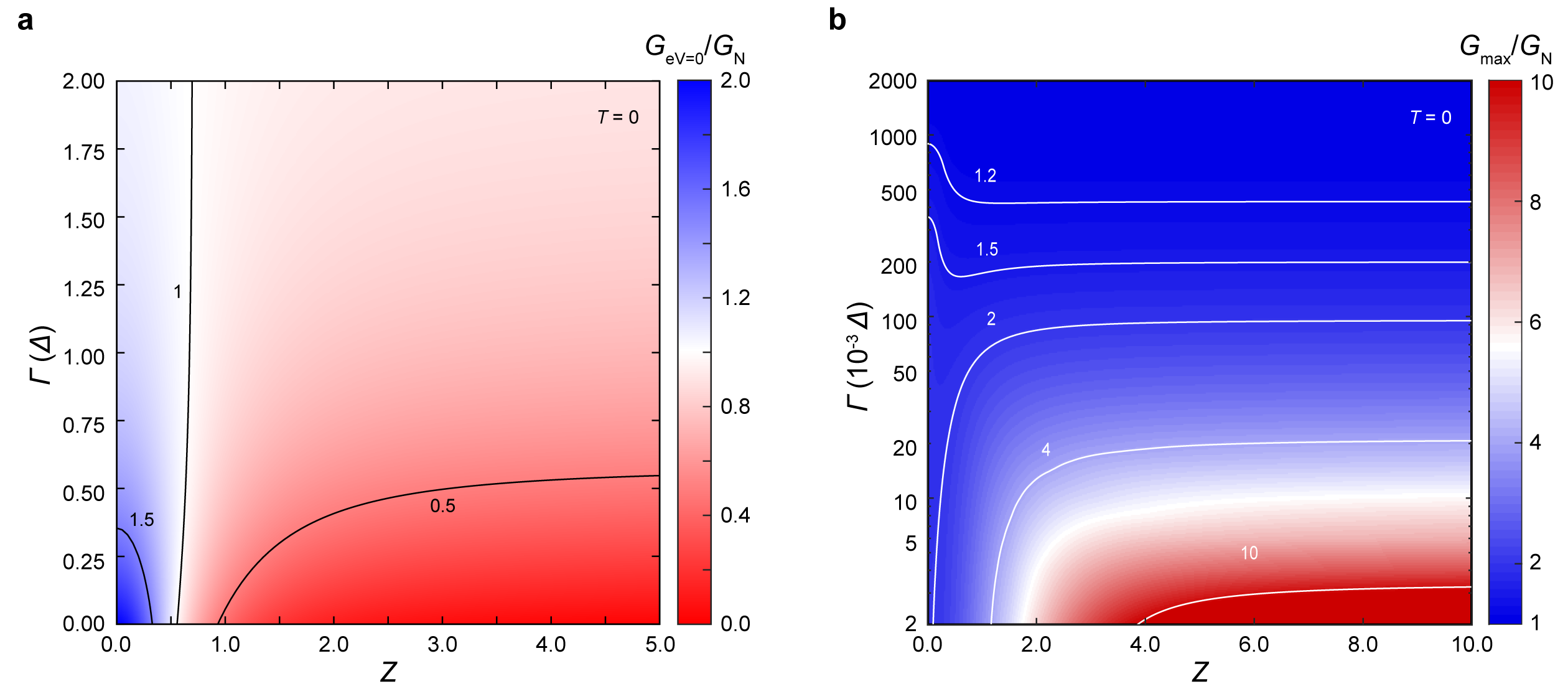}
\caption[Select tunneling spectra parameters versus $Z$ and $\Gamma$]{{\bf Select tunneling spectra parameters versus $Z$ and $\Gamma$.} {\bf a}) Zero-bias conductance ($G_{eV=0}$) normalized by the normal-state conductance ($G_N$) versus $\Gamma$ and $Z$. Contours of $G_{eV=0}/G_N = 1.5$, $G_{eV=0}/G_N = 1.0$, $G_{eV=0}/G_N = 0.5$ are denoted by solid black lines with the corresponding $G_{eV=0}/G_N$ values given. {\bf b}) Maximum spectra conductance ($G_{max}$) normalized by the normal-state conductance ($G_N$) versus $\Gamma$ and $Z$. Contours for various $G_{max}/G_N$ are given by solid white lines with the corresponding $G_{max}/G_N$ values given. All calculations were performed at $T=0$.}
\label{suppfig:ZandGammacontours}
\end{center}
\end{figure*}

\begin{figure*}[ht!]
\begin{center}
\includegraphics[width=7in]{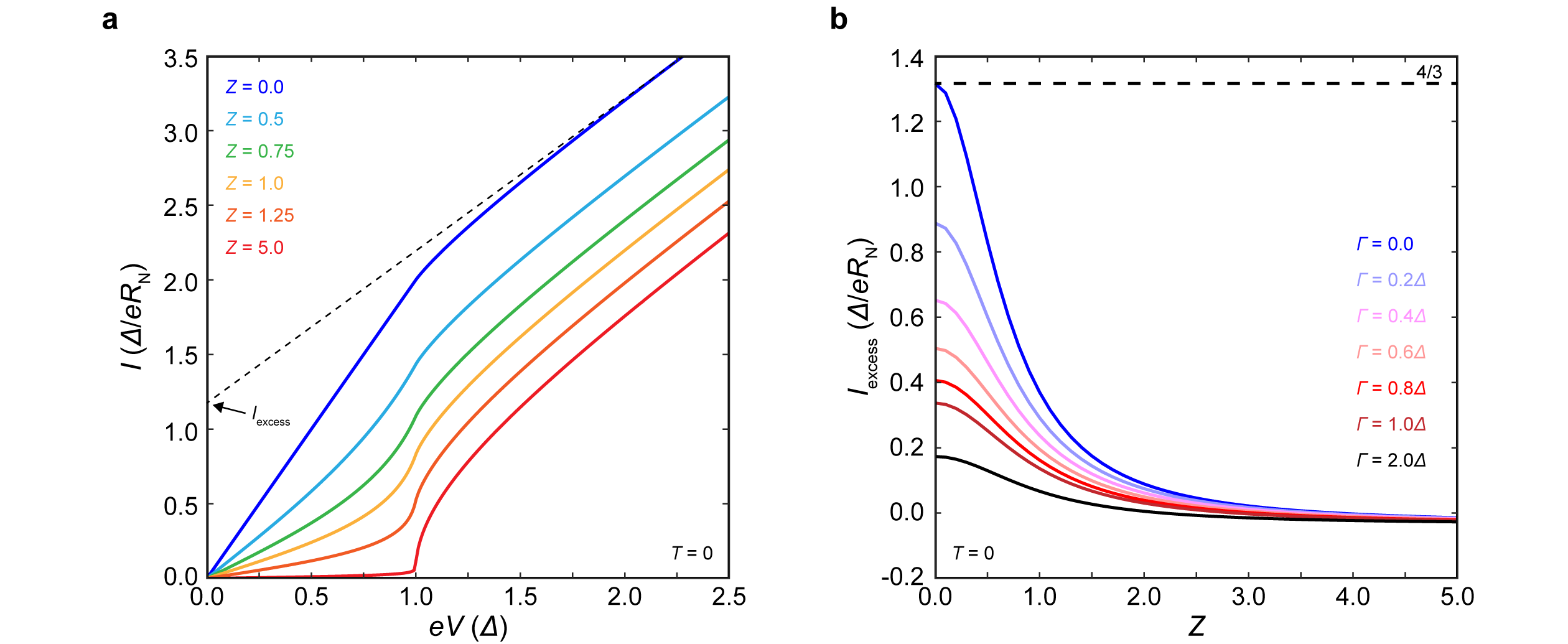}
\caption[Excess current versus $Z$ and $\Gamma$]{{\bf Excess current versus $Z$ and $\Gamma$.} {\bf a}) Calculated $I-V$ curves for various barrier strengths $Z$. The excess current is found by extrapolating the normal-state $I-V$ behavior to zero energy. {\bf b}) Excess current versus barrier strength $Z$ for various $\Gamma$ values (given with respect to the superconducting gap $\Delta$). All calculations were performed at $T=0$.}
\label{suppfig:excesscurrent}
\end{center}
\end{figure*}

In determining the transparency of our measured junctions, it is important to accurately capture the role of $\Gamma$ on the tunnel spectra. A common procedure for extracting $Z$ is to extract the excess current $I_{excess}$ defined as the intersection of the normal-state $I-V$ curves with the $I$ axis (\textbf{Figure \ref{suppfig:excesscurrent}a}). The excess current depends strongly on $Z$ in the low $Z$ regime (\textbf{Figure \ref{suppfig:excesscurrent}b}). However, the $Z$ dependence of the excess current is greatly suppressed with increasing $\Gamma$ (\textbf{Figure \ref{suppfig:excesscurrent}b}). As such, we use a different method to extract $Z$ from our data. The main qualitative role of $\Gamma$ is to suppress sharp features in the spectra (quasiparticle peaks) and push any deviations from the normal-state conductance to the normal-state value (\textbf{Figure \ref{fig:nbse2_NIS}a}). We can accurately extract $Z$ and $\Gamma$ values from our junctions by finding intersecting contours of the zero-bias conductance versus $Z$ and $\Gamma$ (\textbf{Figure \ref{suppfig:ZandGammacontours}a}) and the maximum junction conductance versus $Z$ and $\Gamma$ (\textbf{Figure \ref{suppfig:ZandGammacontours}b}).

\subsection{Simulations of Graphene Tunneling Spectra}
\subsubsection{Tunneling under Zero Magnetic Field}
\begin{figure*}[ht!]
\begin{center}
\includegraphics[width=6.75in]{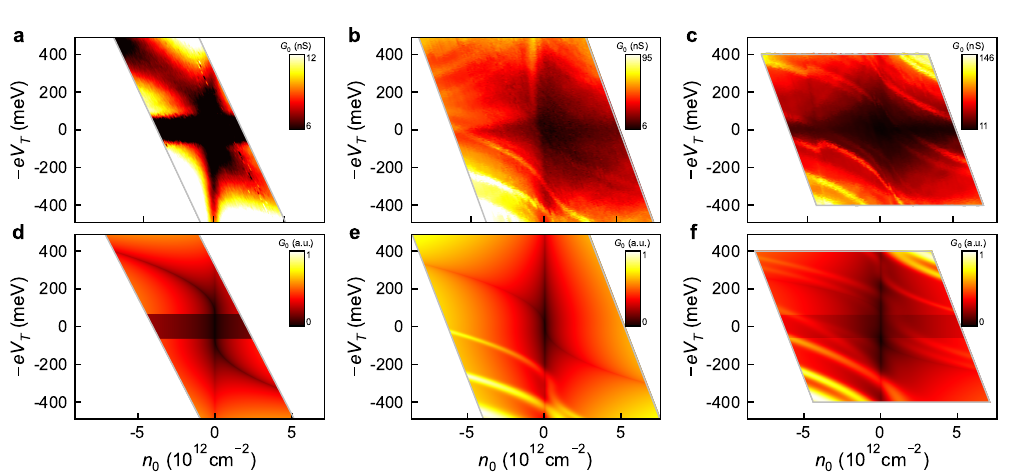}
\caption{\textbf{Measured tunnel spectra for monolayer graphene and simulations under zero magnetic field.} \textbf{a, b, c} Measured dI/dV spectra in $E_T$-$n_0$ space with 7L, 3L and 3L of $h$BN tunnel barriers and 3.08, 0.07 and 0.38 \musq tunnel area, respectively. \textbf{d, e, f} Corresponding simulations of dI/dV with contributions from tunneling, in-plane transport, and phonon gap.
}
\label{suppfig:simulation}
\end{center}
\end{figure*}

We modeled the measured conductance of monolayer graphene based on four contributions: direct tunneling through the via contacts (1 contribution for each contact), tunneling in the presence of the phonon gap in graphene, and in-plane transport. At zero magnetic field, graphene's band structure is a Dirac cone where the location of the charge neutrality point shifts by phonon energy $E_T \sim 63$ meV for phonon-assisted tunneling, and remains unchanged for in-plane transport. We also included several defect states in the graphene band structure to model the bright features in devices with 3L of \textit{h}BN. \textbf{Figure \ref{suppfig:simulation}} shows simulations of conductance for devices with various tunnel barrier thickness and area in comparison with the measured conductance through via contacts. We can see that the phonon-assisted tunneling dominates in the device with 7L of \textit{h}BN, and the bright features in the spectra of 3L devices can be explained by defect states in graphene. Moreover, all devices exhibit features associated with in-plane transport (dip in conductance near $n_0=0$).

\subsubsection{Tunneling under Finite Magnetic Field}
\begin{figure*}[ht!]
\begin{center}
\includegraphics[width=6.75in]{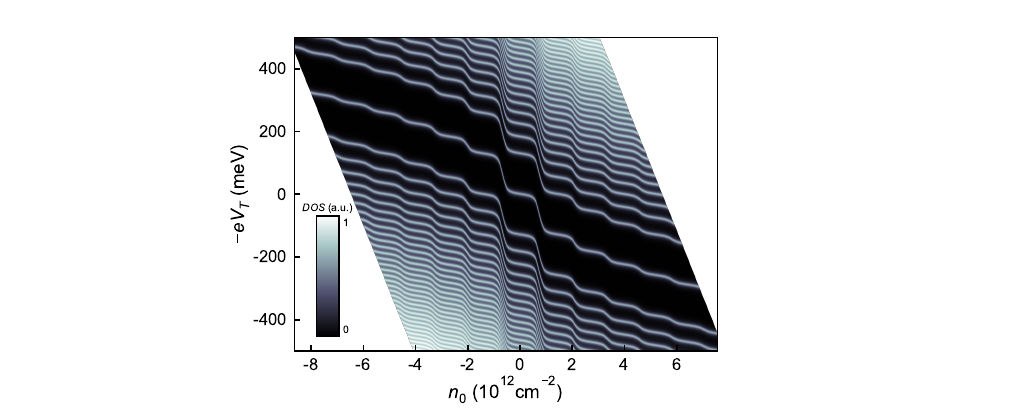}
\caption{\textbf{Self-consistent calculation of DOS in monolayer graphene.} Simulation at B = 12.5T within the experimentally reachable parameter ranges.}
\label{suppfig:LL_simulation}
\end{center}
\end{figure*}

We performed a self-consistent calculation \cite{malec_graphene_tunnel} to model the density of states in graphene with varying doping density and tunneling energy. \textbf{Figure \ref{suppfig:LL_simulation}} shows a representative simulation at B = 12.5T with a similar parameter range as in the experiments.

\clearpage

%%%%%%.   SUPP. FIG. %%%%%%%%%%
% \newpage
\begin{figure*}[ht!]
\begin{center}
\includegraphics[width=3.375in]{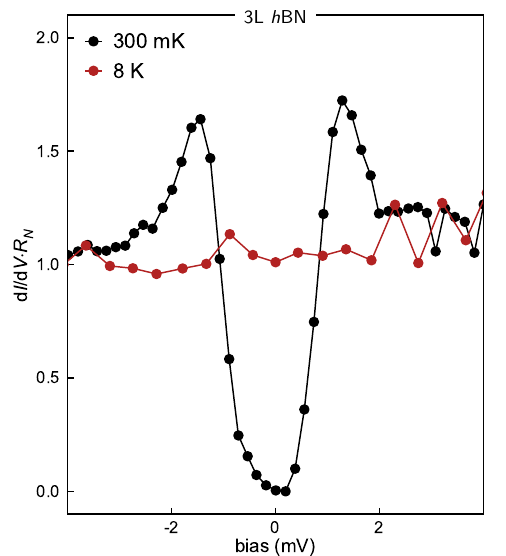}
\caption[NbSe2 3L hBN]{{\bf Via-based tunneling into \nbse with 3L \textit{h}BN.} {Measurements of differential conductance $G = dI/dV_T$ versus junction bias normalized to the normal-state resistance $R_N$ for $T<T_C$ (solid black dots and line) and $T>T_C$ (solid red dots and line) for a 3L \textit{h}BN Au/\textit{h}BN/NbSe\textsubscript{2} tunnel junction. The cross-sectional area of the via tunnel contact is $\sim 0.6$ \musq.}}
\label{suppfig:3LhBNNbSe2}
\end{center}
\end{figure*}
%%%%%%%%%%%%%%%%%%%%%%%%%%%%

%%%%%%.   SUPP. FIG. %%%%%%%%%%
\newpage
%%%%%%%%%%%%%%%%%%%%%%%%%%%%

%%%%%%.   SUPP. FIG. %%%%%%%%%%
\newpage
\begin{figure*}[ht!]
\begin{center}
\includegraphics[width=7in]{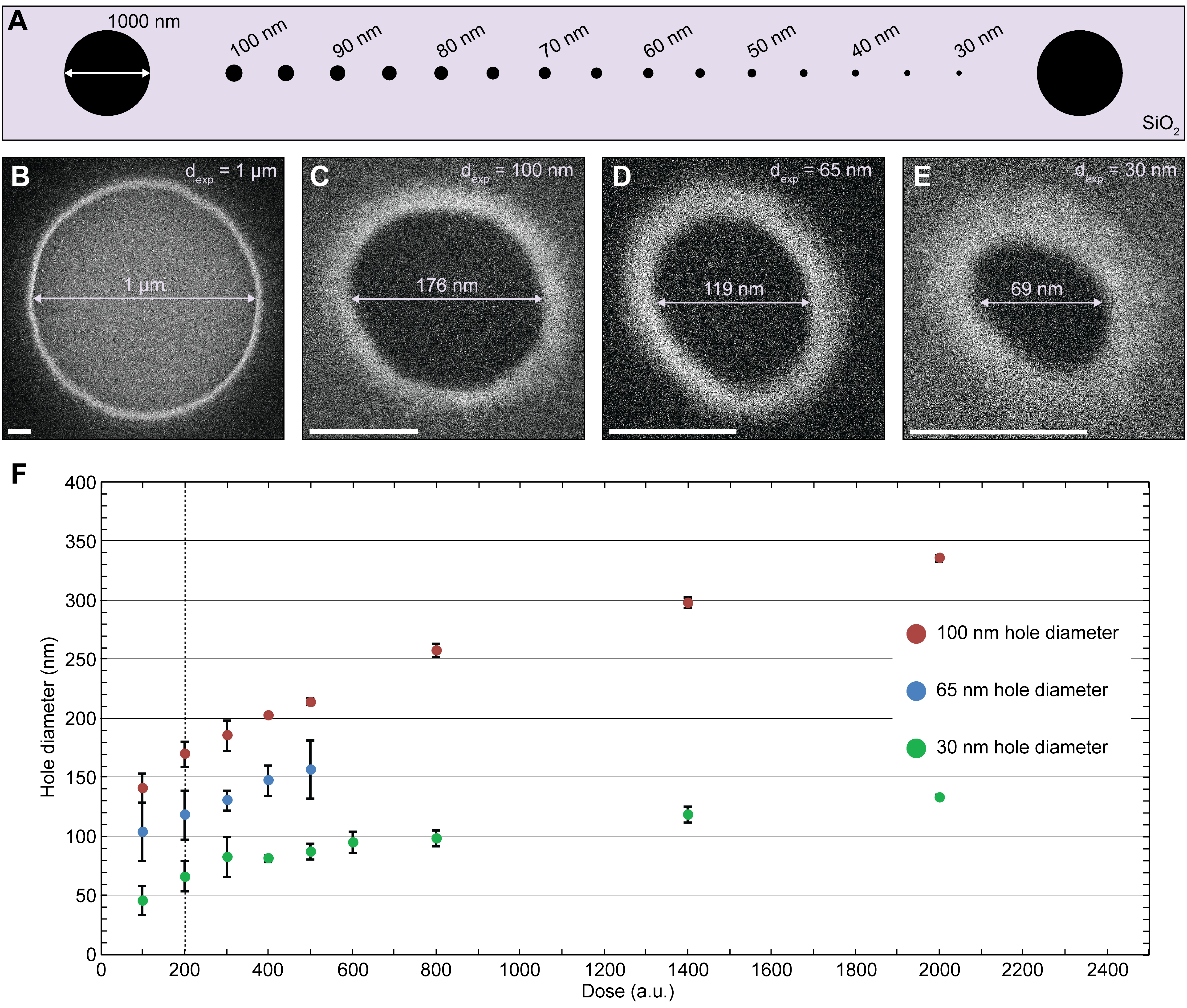}
\caption[Characterization of small-area via contacts]{{\bf Characterization of small-area via contacts.}{ \bf{a}}) Schematic of the lithography design used to dose test the small-area via contacts on SiO\textsubscript{2}. The design contains circles ranging in diameter from 100 nm down to 30 nm in increments of 5 nm. Large 1$\SI{}{\micro m}$ holes are also written for reference. {\bf{b-e}}) Scanning electron microscopy (SEM) images of etched holes in SiO\textsubscript{2} with intended diameters of 1$\SI{}{\micro m}$ ({\bf{b}}), 100 nm ({\bf{c}}), 65 nm ({\bf{d}}), and 30 nm ({\bf{e}}). All scale bars are 100$\SI{}{\micro m}$. {\bf{f}}) Measured hole diameter versus electron dose for 30 nm (solid green dots), 65 nm (solid blue dots), and 100 nm (solid red dots) hole designs. The error bars represent the standard deviation.}
\label{suppfig:small_via_dose}
\end{center}
\end{figure*}
%%%%%%%%%%%%%%%%%%%%%%%%%%%%

%%%%%%.   SUPP. FIG. %%%%%%%%%%
\newpage
\begin{figure*}[ht!]
\begin{center}
\includegraphics[width=7in]{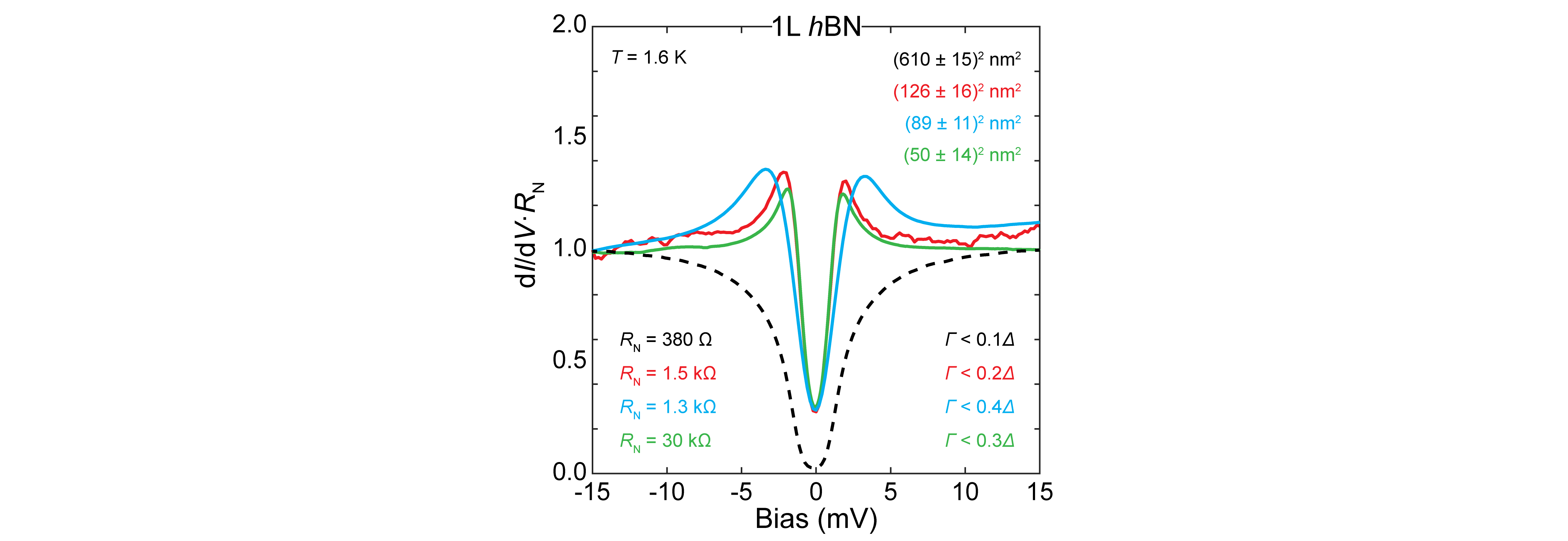}
\caption[Effect of junction size on interface inhomogeneity $\Gamma$ for 1L \textit{h}BN junctions.]{{\bf Effect of junction size on interface inhomogeneity $\Gamma$ for 1L \textit{h}BN junctions.} {Au/\textit{h}BN/NbSe\textsubscript{2} tunnel conductance
versus junction bias normalized to the normal-state conductance for various contact cross-sectional
areas for 1L \textit{h}BN tunnel barriers. Corresponding junction sizes, normal-state resistances, and extracted $\Gamma$ values
are given in the inset.}}
\label{suppfig:1L_areadep}
\end{center}
\end{figure*}
%%%%%%%%%%%%%%%%%%%%%%%%%%%%

%%%%%%.   SUPP. FIG. %%%%%%%%%%
% \newpage
\begin{figure*}[ht!]
\begin{center}
\includegraphics[width=6.75in]{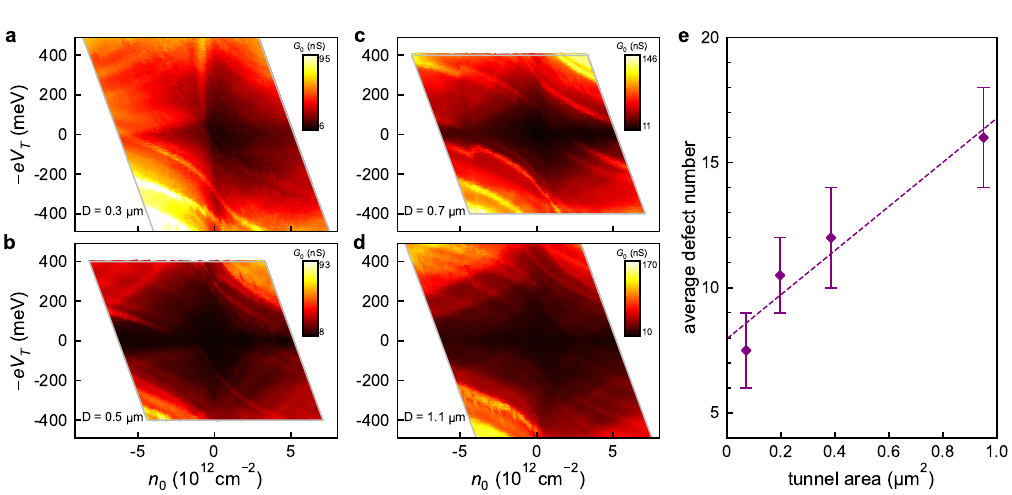}
\caption{\textbf{Number of observed defects in \textit{h}BN tunnel barrier estimated from monolayer graphene tunnel spectra.} \textbf{a-d} Measured dI/dV spectra in $E_T$-$n_0$ space with 3L of $h$BN tunnel barriers and 0.07, 0.20, 0.38 and 0.95 \musq tunnel area, respectively. Diameters of tunnel contacts are given in the insets. \textbf{e} Number of observed defects in (\textbf{a})-(\textbf{d}) versus tunnel area, adapted from Fig.~\ref{fig:defects_temp}\panelb.}
\label{suppfig:areadefects}
\end{center}
\end{figure*}
%%%%%%%%%%%%%%%%%%%%%%%%%%%%

%%%%%%.   SUPP. FIG. %%%%%%%%%%
\newpage
%%%%%%%%%%%%%%%%%%%%%%%%%%%%

%%%%%%.   SUPP. FIG. %%%%%%%%%%
%%%%%%%%%%%%%%%%%%%%%%%%%%%%

%%%%%%.   SUPP. FIG. %%%%%%%%%%
\newpage
%%%%%%%%%%%%%%%%%%%%%%%%%%%%

%%%%%%.   SUPP. FIG. %%%%%%%%%%
\newpage

%%%%%%%%%%%%%%%%%%%%%%%%%%%%

%%%%%%.   SUPP. FIG. %%%%%%%%%%
% \newpage
%%%%%%%%%%%%%%%%%%%%%%%%%%%%

%%%%%%.   SUPP. FIG. %%%%%%%%%%
% \newpage
%%%%%%%%%%%%%%%%%%%%%%%%%%%%

%%%%%%.   SUPP. FIG. %%%%%%%%%%
\newpage
%%%%%%%%%%%%%%%%%%%%%%%%%%%%

%%%%%%.   SUPP. FIG. %%%%%%%%%%
\newpage
\begin{figure*}[ht!]
\begin{center}
\includegraphics[width=7in]{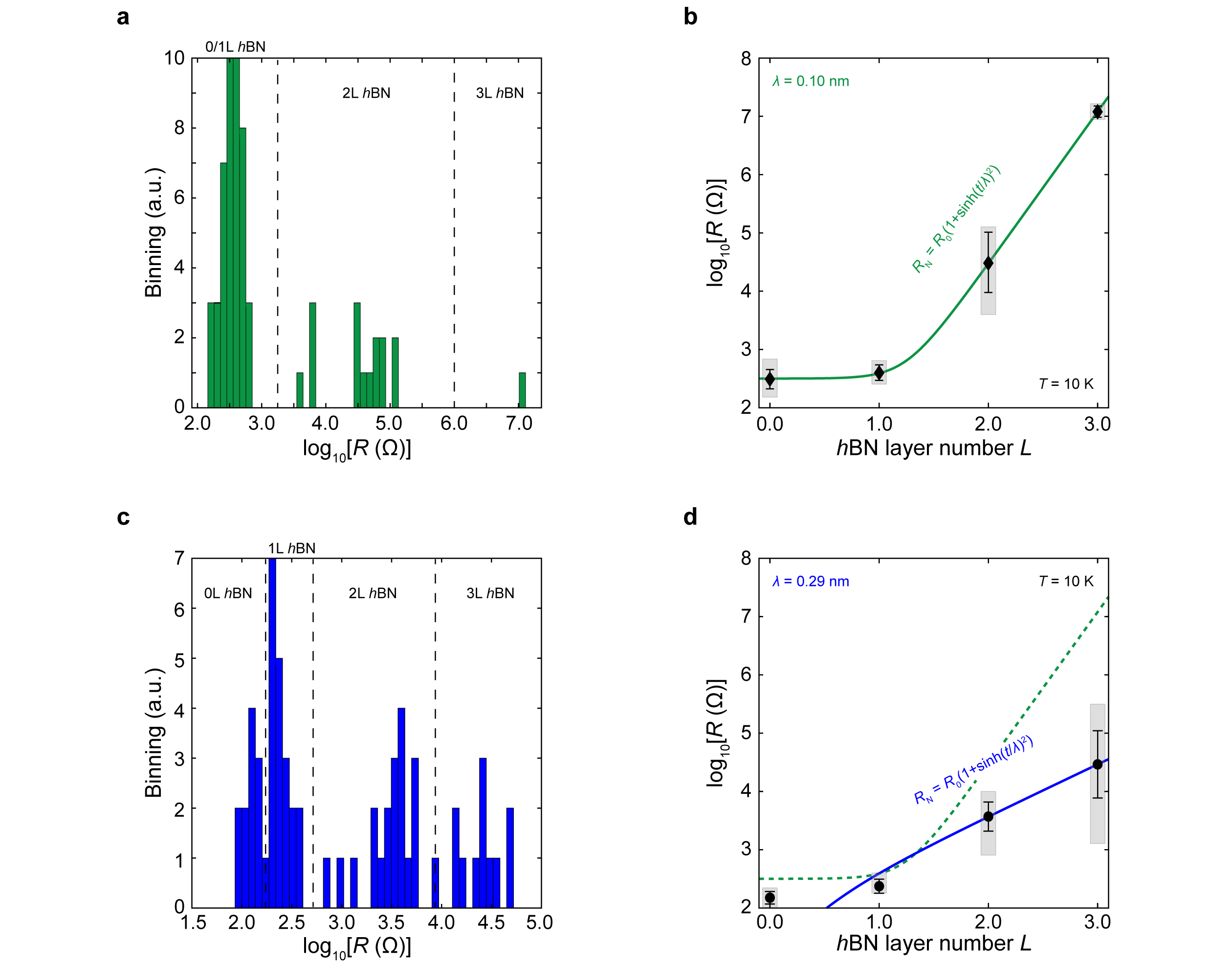}
\caption[Normal-state NIS junction resistance versus \textit{h}BN thickness]{{\bf Normal-state NIS junction resistance versus \textit{h}BN thickness.} {\bf a}, {\bf c}), Histograms of normal-state resistances for various Au/\textit{h}BN/NbSe\textsubscript{2} junctions fabricated with ultra-clean \textit{h}BN flakes ({\bf a}) and intentionally carbon-defected \textit{h}BN flakes ({\bf c}). Resistances were measured at $T=10$ K. Dashed black lines separate the histograms into discrete \textit{h}BN thicknesses. {\bf b}, {\bf d}) Plot of normal-state resistance versus \textit{h}BN thickness for Au/\textit{h}BN/NbSe\textsubscript{2} junctions fabricated with ultra-clean ({\bf b}) and carbon-defected ({\bf d}) \textit{h}BN. Thickness of \textit{h}BN was determined from optical contrast. Solid dark green ({\bf b}) and solid blue ({\bf d}) lines are fits to the data. Extracted fit parameters are given in the insets. Error bars and grey boxes represent the standard deviation and data range for each \textit{h}BN thickness, respectively. In ({\bf d}), the dashed dark green line is the fit from ({\bf b}) for comparison. For all devices, the cross-sectional area of the via tunnel contacts is $\sim 0.6$ \musq.}
\label{suppfig:normal_resistance_vesus_thickness}
\end{center}
\end{figure*}
%%%%%%%%%%%%%%%%%%%%%%%%%%%%

%%%%%%.   SUPP. FIG. %%%%%%%%%%
\newpage
\begin{figure*}[ht!]
\begin{center}
\includegraphics[width=7in]{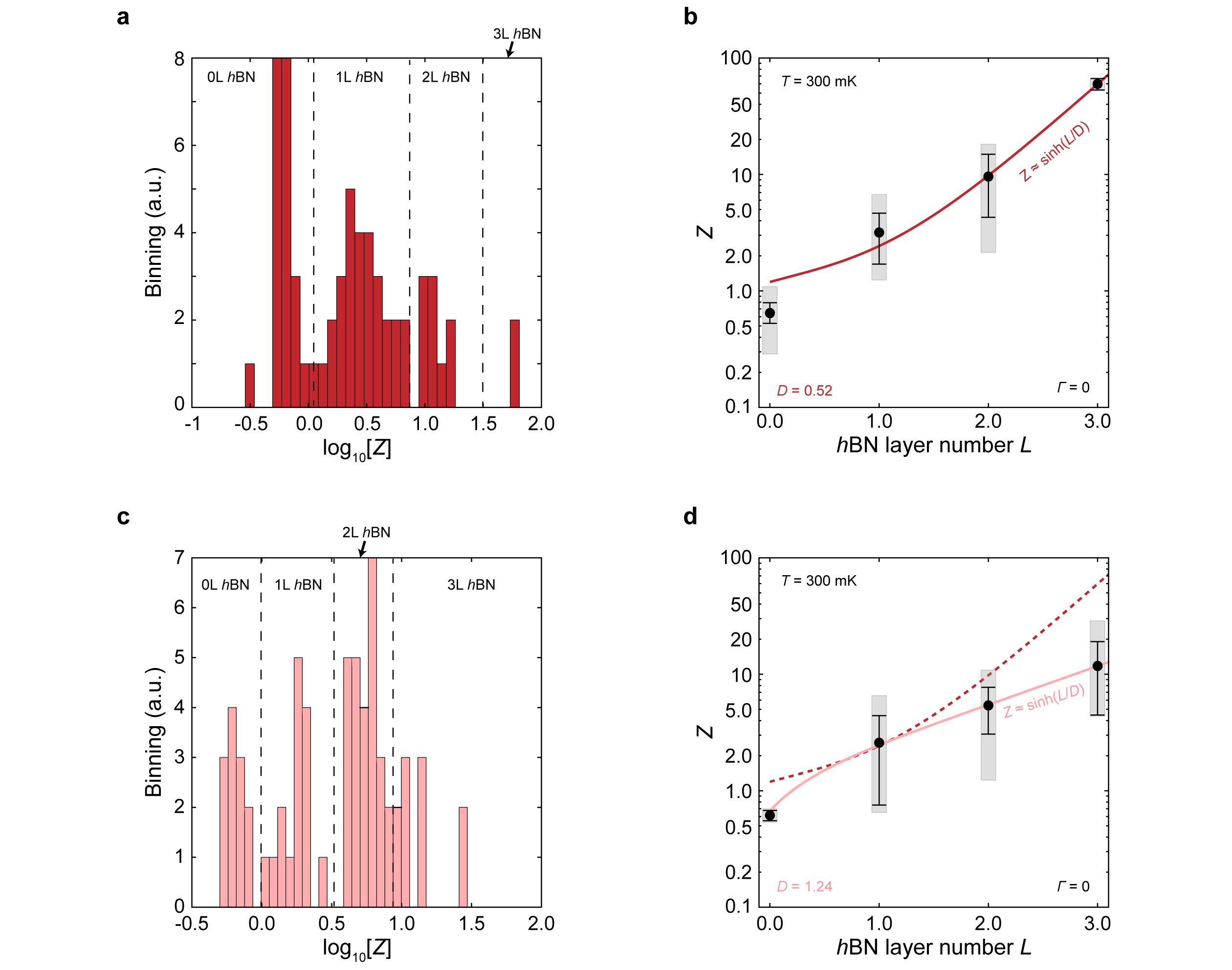}
\caption[NIS junction barrier strength $Z$ versus \textit{h}BN thickness]{{\bf NIS junction barrier strength $Z$ versus \textit{h}BN thickness.} {\bf a}, {\bf c}), Histograms of barrier strength $Z$ for various Au/\textit{h}BN/NbSe\textsubscript{2} junctions fabricated with ultra-clean \textit{h}BN flakes ({\bf a}) and intentionally carbon-defected \textit{h}BN flakes ({\bf c}). $Z$ values were measured at $T=300$ mK. Dashed black lines separate the histograms into discrete \textit{h}BN thicknesses. {\bf b}, {\bf d}) plot of barrier strength $Z$ versus \textit{h}BN thickness for junctions fabricated from ultra-clean ({\bf b}) and carbon-defected ({\bf d}) \textit{h}BN, respectively. \textit{h}BN thicknesses were determined by optical contrast. Solid red ({\bf b}) and pink ({\bf d}) lines are fits to the data. Extracted fit parameters are given in the insets. Error bars and grey boxes represent the standard deviation and data range for each \textit{h}BN thickness, respectively. In ({\bf d}), the dashed red line is the fit to the data in ({\bf b}) for comparison. For all devices, the cross-sectional area of the via tunnel contacts is $\sim 0.6$ \musq.}
\label{suppfig:Z_versus_thickness}
\end{center}
\end{figure*}
%%%%%%%%%%%%%%%%%%%%%%%%%%%%

%%%%%%.   SUPP. FIG. %%%%%%%%%%
\newpage
\begin{figure*}[ht!]
\begin{center}
\includegraphics[width=7in]{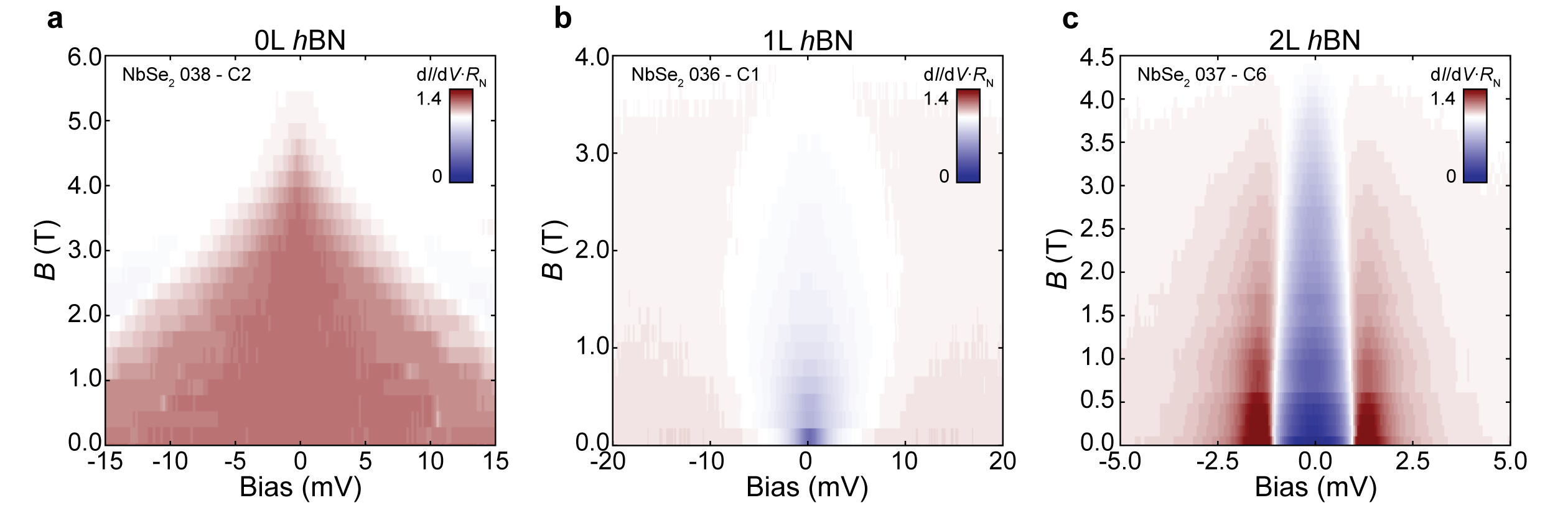}
\caption[Magnetic-field dependence of Au/\textit{h}BN/NbSe\textsubscript{2} tunneling spectra]{{\bf Magnetic-field dependence of Au/\textit{h}BN/NbSe\textsubscript{2} tunneling spectra.} {Contour maps of tunnel conductance
versus junction bias and external magnetic field ({\em B}) for tunnel junctions with 0 ({\bf a}), 1 ({\bf b}), and 2 ({\bf c}) layers of \textit{h}BN as the barrier. Each trace is normalized to the normal-state conductance. All data was acquired at {\em T} $\sim$ 300 mK. For all devices, the cross-sectional area of the via tunnel contacts is $\sim 0.6$ \musq.}}
\label{suppfig:nbse2_field_dependence}
\end{center}
\end{figure*}
%%%%%%%%%%%%%%%%%%%%%%%%%%%%

%%%%%%.   SUPP. FIG. %%%%%%%%%%
% \newpage
\begin{figure*}[ht!]
\begin{center}
\includegraphics[width=7in]{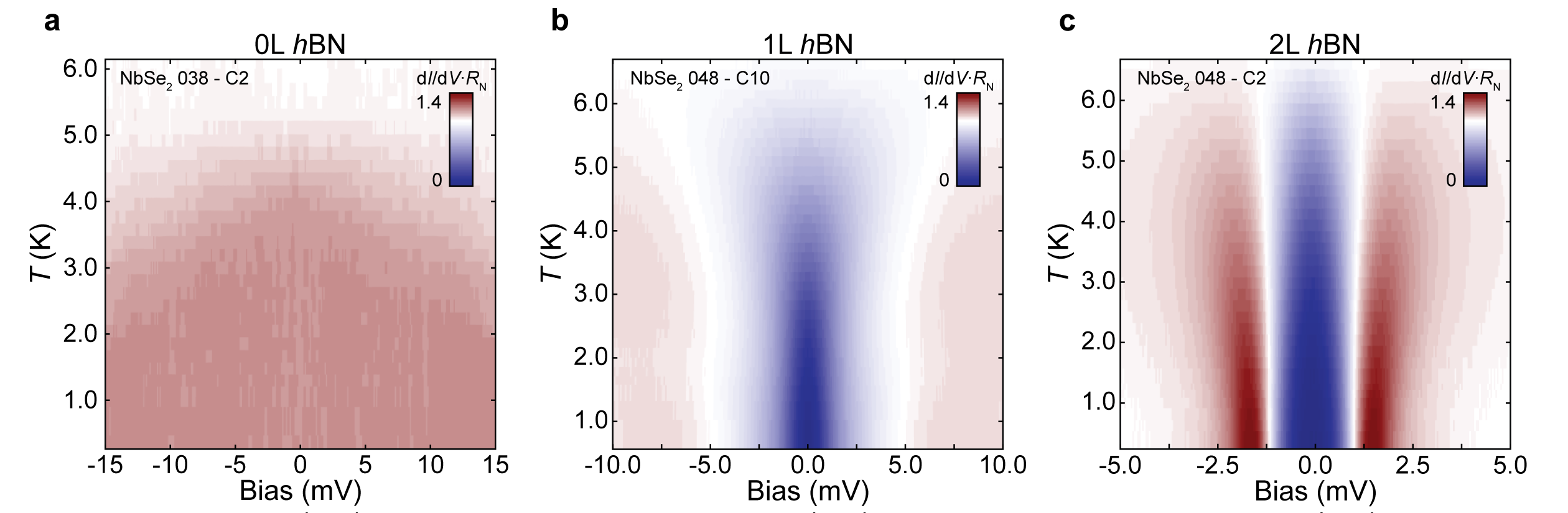}
\caption[Temperature dependence of Au/\textit{h}BN/NbSe\textsubscript{2} tunneling spectra]{{\bf Temperature dependence of Au/\textit{h}BN/NbSe\textsubscript{2} tunneling spectra.} {Contour maps of tunnel conductance
versus junction bias and temperature ({\em T}) for tunnel junctions with 0 ({\bf a}), 1 ({\bf b}), and 2 ({\bf c}) layers of \textit{h}BN as the barrier. Each trace is normalized to the normal-state conductance. All data was acquired at {\em B} = 0 T. For all devices, the cross-sectional area of the via tunnel contacts is $\sim 0.6$ \musq.}}
\label{suppfig:nbse2_temp_dependence}
\end{center}
\end{figure*}
%%%%%%%%%%%%%%%%%%%%%%%%%%%%

% \newpage
% \bibliography{SI_alone}

% \end{document}

% \clearpage
% \section{TOC Graphic}

% %%%%%%.   TOC   %%%%%%%%%%
% \begin{figure}[ht!]
% \begin{center}
% \includegraphics[width=3.25in]{figures/TOC_graphic/TOC_graphic.pdf}
% \label{fig:TOC}
% \end{center}
% \end{figure}
% %%%%%%%%%%%%%%%%%%%%%%%%%%%%

\end{document}